%% file: main.tex
\documentclass[preprint,12pt]{elsarticle}

\usepackage[english]{babel}

\usepackage[letterpaper,top=2cm,bottom=2cm,left=3cm,right=3cm,marginparwidth=1.75cm]{geometry}

\usepackage[colorlinks=true, allcolors=blue]{hyperref}
\usepackage{amsmath} 
\usepackage{verbatim}
\usepackage{subcaption}
\graphicspath{ {./images/} }

\usepackage{algorithm,algorithmic}

\usepackage{comment}
\usepackage{enumitem} 
\usepackage{array}
\usepackage{mathtools} 
\usepackage{adjustbox}
\usepackage{tabularx,multicol,multirow,booktabs}
\usepackage{rotating}
\usepackage{multirow}
\usepackage{tabularx}
\usepackage{xcolor}
\usepackage{colortbl}

\usepackage{amsfonts} 
\usepackage{amssymb} 
\usepackage{graphicx} 

\usepackage{lineno} 

\makeatletter
\def\ps@pprintTitle{%
 \let\@oddhead\@empty
 \let\@evenhead\@empty
 \def\@oddfoot{}%
 \let\@evenfoot\@oddfoot}
\makeatother

\begin{document}
\extrafloats{100}
\begin{frontmatter}


\title{Integrating acoustic tapping with a UAV platform for tile condition classification}



\author[unm1]{Piedad J. Miranda}
\author[unm1]{Ronan Reza}
\author[fiu1]{Leonel Lagos}
\author[fiu2]{Mackenson Telusma}
\author[srnl1]{Christine A. Langton}
\author[unm1]{Fernando Moreu\corref{cor1}}

\cortext[cor1]{Corresponding author: fmoreu@unm.edu (F. Moreu)}

\address[unm1]{Department of Civil, Construction and Environmental Engineering, University of New Mexico, 210 University Blvd NE, Albuquerque, NM 87131}
\address[fiu1]{Moss Department of Construction Management, Florida International University, 10555 West Flagler Street, EC 2900, Miami, FL 33174}
\address[fiu2]{Applied Research Center, Florida International University, 10555 West Flagler Street, EC 2100, Miami, FL 33174}
\address[srnl1]{Savannah River National laboratory, Aiken, SC 29808}

\date{} 

\begin{abstract}

Ensuring the structural integrity of building tiles is important for public safety and the durability of urban infrastructure. This study proposes a controlled experimental framework to quantify the effect of Unmanned Aerial vehicle (UAV) induced dynamic perturbations on acoustic tap-testing reliability for facade inspection. 
This work explicitly analyzes vibration-induced degradation and introduces an energy-based signal correction method to preserve classification performance under motion disturbances. In addition, Principal Component Analysis (PCA) is applied to process and classify wirelessly acquired acoustic data, reducing dimensionality while preserving key defect-related features. 
A Stewart platform is used to reproduce controlled oscillatory conditions derived from UAV flight characterization, enabling systematic evaluation across multiple vibration amplitudes. Results show that classification accuracy degrades significantly under increasing perturbations, but can be restored above 98\% using the proposed energy-based filtering approach.

\end{abstract}

\end{frontmatter}

\input{Sec1}

\input{Sec2}

\input{Sec2_5}

\input{Sec4}

\input{Sec3}

\input{Sec6}

\input{Sec7}

\bibliographystyle{IEEEtran}
\bibliography{sample}

\end{document}

%% file: Sec1.tex

\section{Introduction}



The deterioration of building tiles over time present safety concerns, leading to the detachment of deteriorated components that can fall off without warning. 
While inspection protocols for early defect detection exist, they vary significantly in scope and frequency, and face inherent limitations in scalability, cost-effectiveness, and inspector safety \cite{adhikari2023usage}. Current visual inspections, aim to assess structural conditions and identify defects that can compromise the integrity of the building \cite{hassani2023systematic}. However, visual approaches prove inadequate to detecting subsurface defects or quantifying material degradation with sufficient resolution \cite{rakha2018review}.


\begin{figure}[h]
\centering
\begin{subfigure}[b]{0.295\linewidth}
\includegraphics[width=0.98\textwidth]{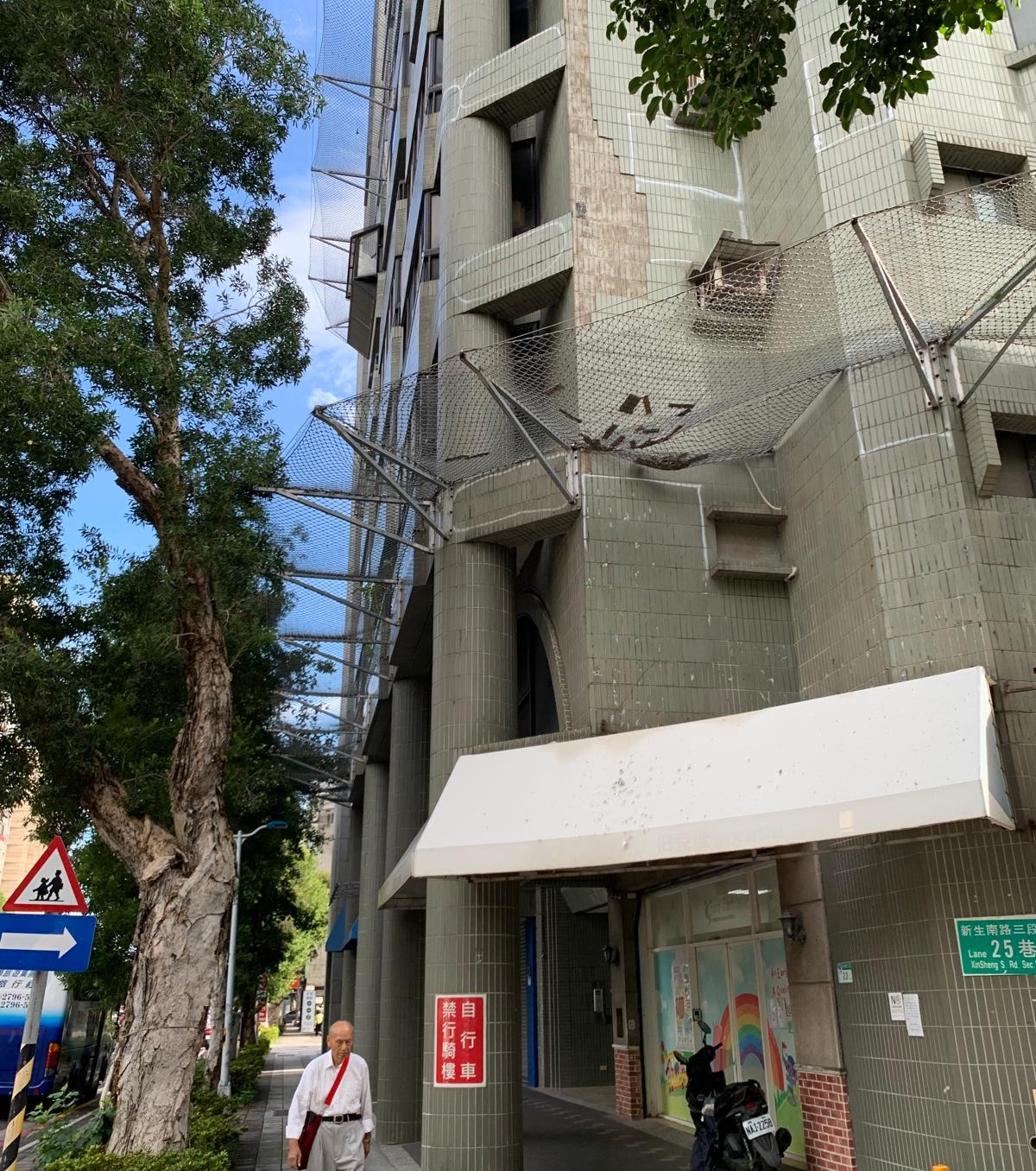}
\subcaption{}\label{fig:atile} \end{subfigure}
\begin{subfigure}[b]{0.45\linewidth}
\includegraphics[width=0.98\textwidth]{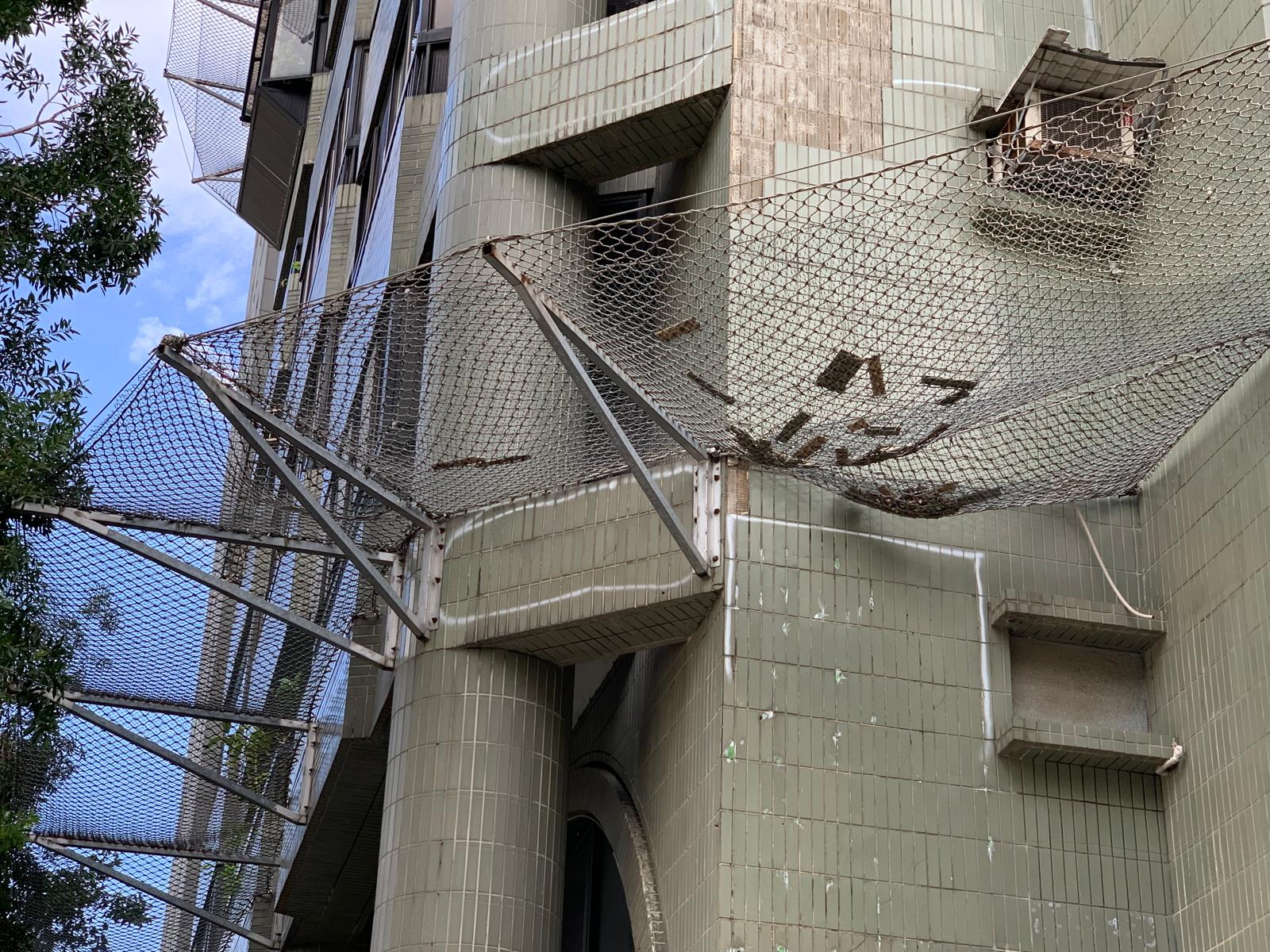}
\subcaption{}\label{fig:btile} \end{subfigure}
\caption{Deterioration of tiles in buildings due to the passage of time and environmental conditions.} \label{fig:tile_problem}
\end{figure}


The integration of computer vision (CV) techniques has recently transformed traditional inspection methods. Advanced algorithms, including Machine Learning (ML) and Deep Learning (DL), now enable automated damage detection with enhanced precision and efficiency \cite{silva2024automated}. These approaches are particularly valuable in conservation, as demonstrated by Karimi et al. \cite{karimi2024deep,karimi2024automated}, who combined automated vision systems with thermogravimetry (TGA) and X-ray diffraction (XRD) for comprehensive tile damage assessment in Portuguese cultural heritage buildings. 
Mayya and Alkayenm \cite{mayya2025triple} proposed a method for detection and classification of cracks in historic buildings, bridges and other masonry structures with YOLO-ensemble and classification with K-means. Dong et al. \cite{dong2025new} proposed automated defect detection of complex textured ceramic tiles based on image preprocessing, feature extraction and morphological filtering. Laofor et al. \cite{laofor2012defect} introduced Digital Image Processing (DIP) to detect and quantify defects in tiles. 
However, the use of AI faces some challenges. The first is the need for experts to develop image recognition algorithms and then train the networks with the classes they want to identify in the images. The second challenge is related to the large amount of data needed to train the algorithms. And finally, some limitations of image processing are related to image acquisition, the appearance of the coating, the distance and position of the camera, and to the light conditions at the time of inspection \cite{dias2021critical}. These factors can affect the captured image and cause inaccuracies in defect detection.


The diagnosis of building using drones offers significant advantages in terms of speed, especially in façade assessments. Chen et al. \cite{chen2019opportunities} has developed a structured system for anomaly detection and management based on images captured by drones, allowing the accurate identification of surface damage and cracks in vertical building elements. Lin et al. \cite{lin2025true} proposed an approach where UAVs capture RGB and thermal images. the RGB images generate high quality 3D point clouds detecting fractures and thermal leakage. Zhang et al. \cite{zhang2024integrated} proposed an UAV imagery collection, anomaly detection using ResNet-18, and damage quantification using a Damage Index (DI) to prioritize repairs for roof damage detection. The objective of Ruiz et al. \cite{ruiz2021unmanned} was to perform Digital Image Processing (DIP) analysis for the automatic detection of cracks in ceramic tiles in buildings, associated with UAVs. Peng et al. \cite{peng2017unmanned} proposed a methodology for the quantification of façade damage through the use of cameras mounted on UAV platforms. The developed system allows the systematic acquisition of georeferenced images that, through a semi-automated process of digital photogrammetry and image analysis, generate a detailed map of surface pathologies.
Through the use of drones and extensive information, the system achieves a high degree of accuracy in damage classification~\cite{falorca2021facade}. However, its application is limited to the specific building, which could affect the generalization of the model, thus restricting the machine learning process and, consequently, impairing detection capabilities.




Visual inspection, whether performed manually or through image-based methods, remains limited to surface-level assessment and is influenced by lighting conditions, accessibility, and inspector subjectivity. These limitations have driven the development of automated inspection systems within construction engineering, where repeatability, and operational safety are essential. 
In this context, UAV-based platforms have emerged as a promising solution for facade inspection. However, most UAV-based approaches rely predominantly on visual sensing modalities, inheriting their intrinsic limitations while also being affected by environmental and operational variability. Acoustic tap testing provides a complementary approach capable of detecting internal defects, but its integration with UAV systems introduces dynamic perturbations that affect measurement reliability. The influence of UAV-induced vibrations on acoustic signal integrity and classification performance remains largely unquantified, representing a critical gap in current automated inspection methodologies.

\begin{figure}[!htbp]
    \centering
    \includegraphics[width=0.98\linewidth]{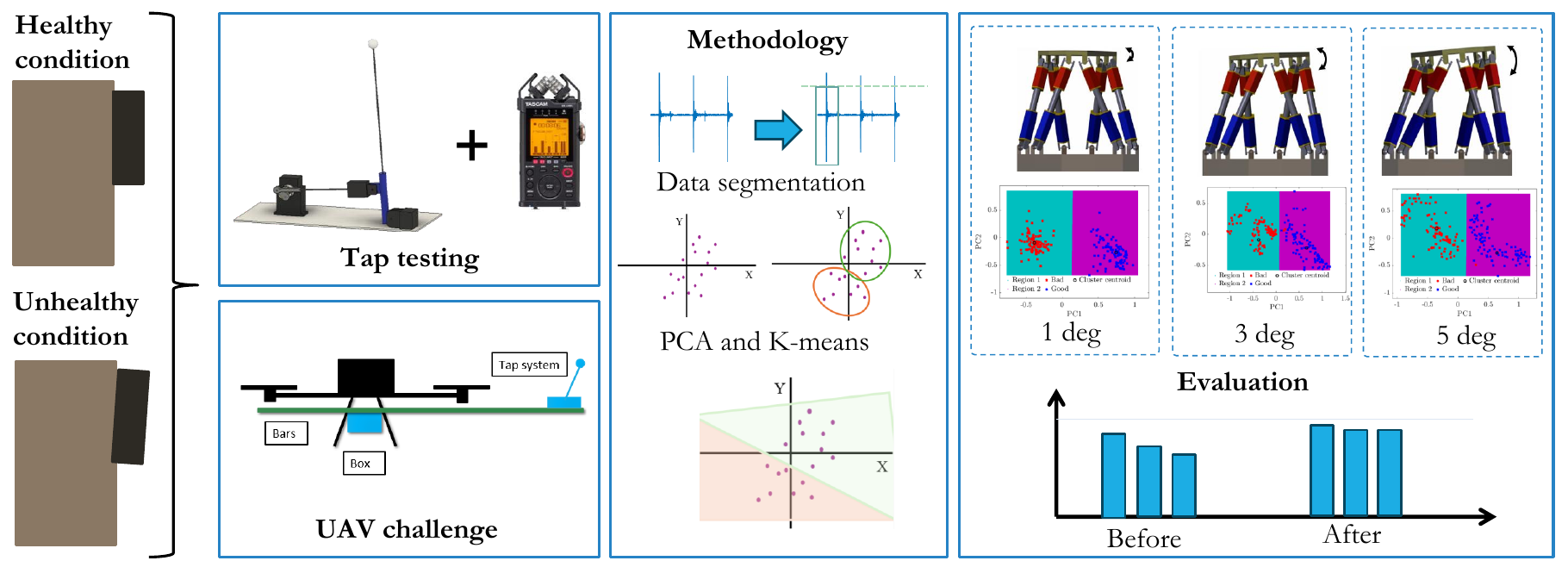}
    \caption{Tap testing summary}
    \label{fig:summary}
\end{figure}

To address this gap, this study proposes a controlled disturbance framework that isolates vibration effects and introduces an energy-based correction methodology to enhance classification robustness under dynamic conditions. 
As shown in Figure~\ref{fig:summary}, we propose a reproducible framework for quantifying the impact of dynamic perturbations on the accuracy of acoustic measurements using the tap testing method \cite{moreu2018remote}. A characterization of the drone behavior is performed using Vicon cameras to capture the flight dynamics. Furthermore, we use the UAV platform to emulate its behavior in various scenarios, allowing us to fine-tune and optimize its performance in detail prior to carrying out activities in the field. 


\section{Background}

Recent approaches in robotics and biomimetic technologies have led to advances in surface inspection, especially with regard to the detection of structural damage through controlled impact and acoustic analysis methods. Nishimura et al. \cite{nishimura2022automated} developed a propeller-type wall-climbing robot designed for hammer inspection. Later, Nishimura et al. \cite{nishimura2024propeller} incorporated a camera, extending the robot's capabilities to capture images and examine hammer hits. Wall-climbing robots are ideal for inspecting the concrete surface, as they can adhere to the surface while maintaining a fixed distance and a constant force, in addition to reducing weight and including mechanisms that ensure a uniform impact force \cite{sukvichai2017design}.
In another instance, Nemati and Dehghan-Niri \cite{nemati2024bio, nemati2023biomimetic} proposed a biometric approach that seeks to replicate the aye-aye testing system, creating repeatable mechanical impacts to detect any changes in its properties. Further, Nemati et al. \cite{nemati2025investigating} integrated the measurement of temperature on the surface of the pinna, during tapping episodes, in order to discern any thermal patterns associated with this percussive behavior \cite{masurkar2024enhancing}. This demonstrated an improvement in sensitivity to detect the condition of the specimens, wood blocks. By other hand, Song et al. \cite{song2024tunnel} planned the generation of impact and the recording of the resulting acoustic echo signal, the pulse-echo method in tunnel lining inspection, complemented by spectral and waveform analysis of the acoustic signals. In Ichikawa et al. \cite{ichikawa2017uav} developed a hammering device that can be mounted on UAV, the device has a microphone to record the sound and find defects at a depth of about 100 mm from the surface. Shoda et al. \cite{shoda2024defect} proposed an ego-noise reduction method based on propeller vibrations, which is beneficial for hammering inspections, where the variability in hammering acoustic sounds is caused by different external conditions. Huang et al. \cite{huang2023development} proposed a method for simultaneous identification of surface, internal and composite (consisting of surface and internal damage) damage in a complex post-disaster environment based on variable frequency hammering. These advances demonstrate the growing potential of integrating multiple technologies to optimize material damage detection, particularly in difficult or previously unexplored conditions. However, there is still a lack of industrial applications that can effectively operate in challenging environments, such as those required for facade inspections today.


On the other hand, hexapod platforms, recognized for their ability to generate and simulated movements. These systems are used to replicate complex motion patterns that can simulate environmental conditions, dynamic loads, and deformation scenarios that structures may encounter Martonka and Fliege \cite{martonka2014hexapod}. Expanding upon one of the uses, they are also used for vibration based damage detection Avci et al. \cite{avci2021review}. By combining ML and sensor data gathered from hexapod testing, they were able to reproduce a range of modes and loading conditions, which further enhances the accuracy of data collection and interpretation. Szabo et al. \cite{burkus2022mechanical} used a hexapod to apply cyclic loads to structures, which are designed to mimic the wear and tear that structures have to endure over time. In Min et al. \cite{min2021port}, they proposed a way to use the hexapods extremely accurate displacement mechanism in order to evaluate structural movement in port infrastructure. Force interaction modeling is another area where Stewart platforms are useful. In Li et al. \cite{zhang2017force}, they have methods for incorporating force sensing systems into parallel platforms to make sure that they have a precise simulation of a real-world stress environment. These platforms also serve as an inspection tool, Vivaldini et al. \cite{teixeira2021intelligent} demonstrated how a platform with DL equipped inspection systems can find minute defects in structural components in an aerospace context. In Bodie et al. \cite{bodie2019omnidirectional}, they used Stewart platforms to test structural components in confined spaces where traditional inspection methods are difficult to apply. 
Its integration with advanced computational techniques makes it a fundamental tool for advancing reliability and safety in aerospace, civil infrastructure and confined space inspections.

Based on signal processing techniques applied to ecoacoustics, authors such as 
Wang et al. \cite{wang2025road} have demonstrated how the combination of acoustic indices, PSD, and DL models (CNN/ResNet50) allows for the assessment of the impact of roads on landscapes by identifying changes in acoustic composition and diversity on a large scale in natural ecosystems. On the other hand, Sánchez-Gendriz \cite{sanchez2021signal} demonstrated how techniques such as Short-Time Fourier Transform (STFT), Power Spectral Density (PSD), and Parseval's theorem can be used to detect and characterize acoustic events in long-term recordings, even in environments with significant background noise. Sánchez-Gendriz et al. \cite{sanchez2024exploring} demonstrated how calculating energy in specific frequency bands from PSD, combined with PCA, allows for the efficient characterization of the temporal and spectral structure of fish choruses in underwater acoustic monitoring, facilitating the extraction of dominant patterns and the visualization of soundscape dynamics. This approach allows for the identification of temporal-spectral patterns, offering a robust methodology for the automatic analysis of environmental or biological sounds in acoustic monitoring applications.



%% file: Sec2.tex

\section{Tap testing device}

\subsection{Hammer}
The tap testing hammer is composed of a sphere, which is connected to a mechanism by means of a flexible beam of 1/8'' in diameter and 23'' in length. 
The design of the hammer allows it to strike the surface of the object under test as many times as required, ensuring that each impact is made with consistent energy, direction and frequency. 
This consistency is made possible by the automated control of the device, which minimizes human error that normally arises due to variability in the force, angle or rate of each tap when inspections are performed manually. By eliminating human subjectivity, the device ensures that each hit is virtually identical, making it easier to compare data and identify changes or damage to the structure over time.

\begin{figure}[!ht]
\centering
\begin{subfigure}[b]{0.52\linewidth}
\includegraphics[width=0.98\textwidth]{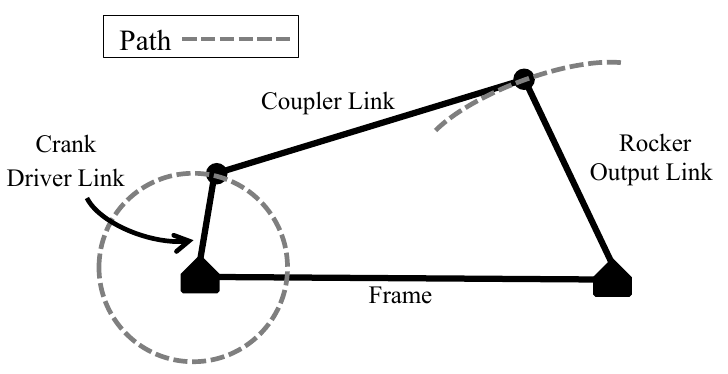}
\label{fig:a} \subcaption{}
\end{subfigure}
\begin{subfigure}[b]{0.43\linewidth}
\includegraphics[width=0.98\textwidth]{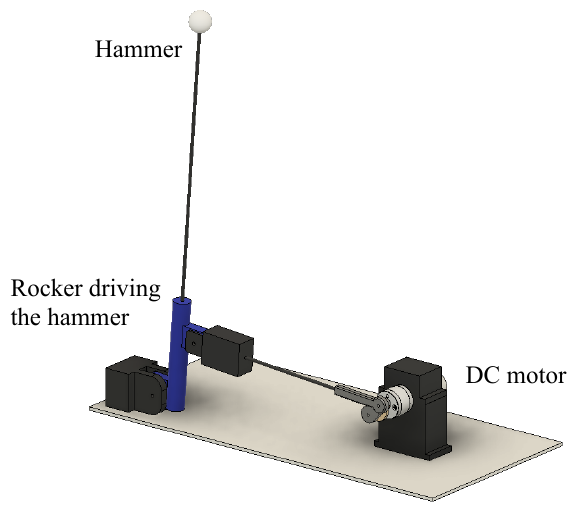}
\label{fig:b} \subcaption{}
\end{subfigure}
\caption{Impact mechanism: (a) 
four-bar linkage concept; (b) tap testing device}
\label{fig:entrenamientopruebas}
\end{figure}

\subsection{Tapping mechanism}

The tap testing mechanism is designed to replicate the manual tapping motion performed by inspectors during field assessments of structural integrity. This system employs a crank-rocker mechanism to convert rotational motion into a controlled, reciprocating strike, simulating the precise impact of an inspector's hammer \cite{nasimi2022use}. The striker hammer head, driven by this mechanism, delivers consistent and measurable impulses to the test surface, enabling the detection of subsurface anomalies such as delamination or voids.

\begin{figure}[htbp]
    \centering
    \includegraphics[width=0.76\linewidth]{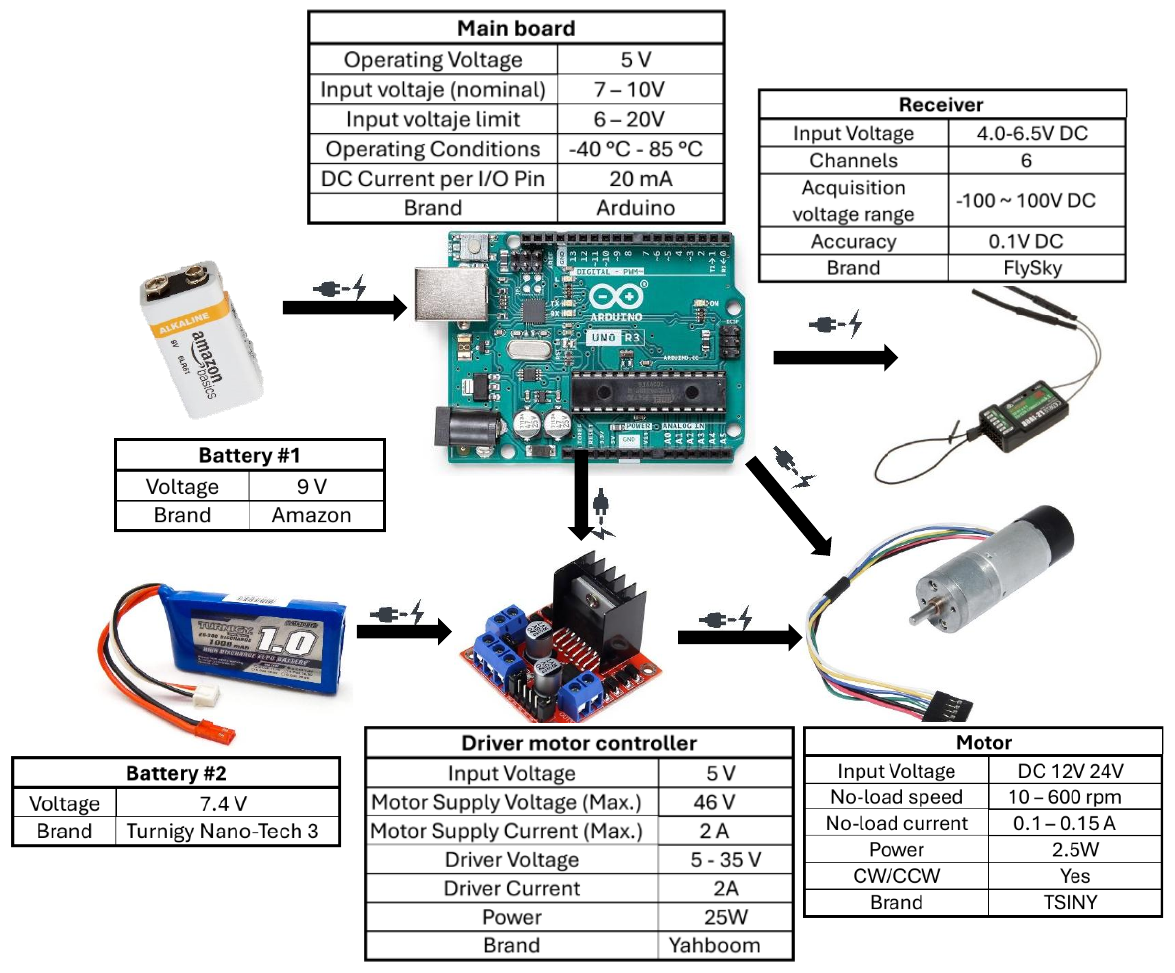}
    \caption{Tap testing circuit components and their specifications.}
    \label{fig:circuit}
\end{figure}

Figure~\ref{fig:circuit} illustrates the schematic representation of the electronic control system governing the tap testing approach. The circuit integrates a microcontroller (Arduino-based main board) to coordinate actuation signals, a motor controller to regulate the striker's motion, and a receiver for wireless command transmission \cite{FlyskyRemote}. The DC motor provides the necessary torque to drive the crank-rocker assembly, while power is supplied through two battery to ensure stable operation. This setup ensures repeatability in strike force and frequency, critical for quantitative defect characterization.

\begin{table}[ht]
    \centering 
    \footnotesize
    \caption{Circuit battery consumption.}
    \begin{tabular} {|l|m{1.6cm}|m{2.3cm}|m{1.6cm}|m{3.5cm}|}
    \hline
    \textbf{Scenario} & \textbf{Battery} & \textbf{Consumption (mA)} & \textbf{Capacity (mAh)} & \textbf{Approximate Autonomy (hours)} \\
        \hline  
        Arduino/Receptor & 9 V \newline 500 mAh & 130 & 500 & 3.85 \\
        \hline
        Motor/Controller & 7.4 V \newline 1000 mAh & 250 & 1000 & 4  \\
        \hline
    \end{tabular}
    \label{tab:autom}
\end{table}

The energy autonomy of the circuit is ensured through a double battery, with power consumption optimized for extended field operation. As detailed in Table~\ref{tab:autom}, the 9 V battery (500 mAh capacity) supplies the Arduino and receiver subsystem, delivering approximately 3.85 hours of continuous operation at a measured consumption rate of 130 mA. Simultaneously, the 7.4 V battery (1000 mAh capacity) powers the motor and controller assembly, sustaining 4 hours of operation under a 250 mA load. This balanced power distribution enables sufficient testing durations for practical field inspections while maintaining portability and energy efficiency

\subsection{ Data Acquisition System} 
\begin{figure}[ht]
    \centering
    \includegraphics[width=0.6\linewidth]{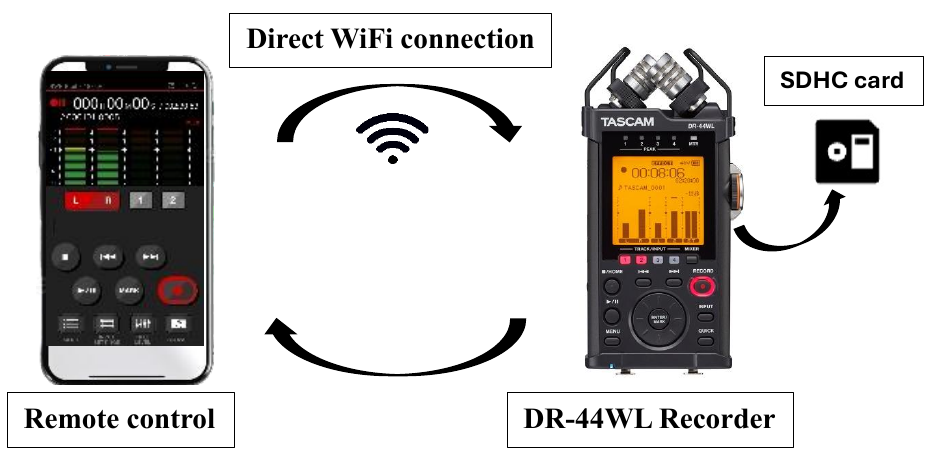}
    \caption{Wireless communication}
    \label{fig:mic1}
\end{figure}

The system has a data acquisition device, which must be controlled remotely \cite{mason2016tap}. Figure~\ref{fig:mic1} shows the wireless communication between a mobile device and a DR-44WL audio recorder via a direct WiFi connection \cite{microphonePCM}. The phone functions as a remote control through the Tascam PCM Recorder application, allowing recording management, such as starting, pausing and adjusting parameters. The DR-44WL stores audio files in its internal memory or on an SD card, without the need for a computer.

The microphone's wireless capabilities
, operate at a frequency of 2.4 GHz under the IEEE 802.11 b/g/n standard. This frequency band is known to provide moderate and reliable signal range in open spaces. 
In addition, the microphone uses a Simple Access Point communication mode, which allows it to connect directly to a device, without the need for an intermediary router. This direct connection simplifies setup and improves portability, making it an effective solution for applications where ease of use and speed of deployment are a priority.

%% file: Sec2_5.tex

\section{UAV characterization}

\subsection{Physical Characteristics}

\begin{figure}[htbp]
    \centering
    \begin{subfigure}[b]{0.355\linewidth}
    \includegraphics[width=0.98\textwidth]{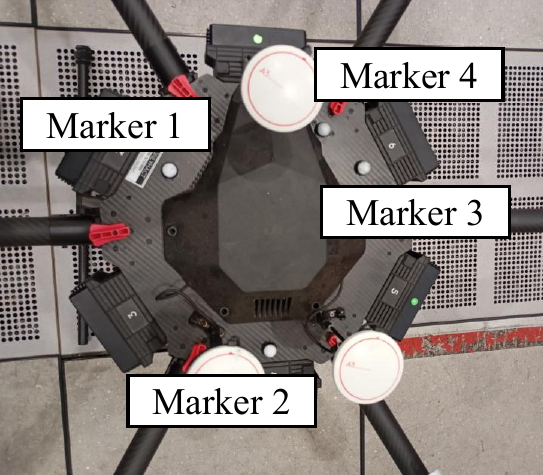}
    \label{fig:charecteristicsa} \end{subfigure}
    \caption{Location of markers on the UAV chassis} \label{fig:tackingUAV1}
\end{figure}

The dynamic characterization of the drone Matrice 600 Pro \cite{dronemodel} was performed using a Vicon Valkyrie VK8 motion capture system \cite{viconmodel}, configured with an array of 10 cameras 
and with a sampling rate of 100 Hz. Vibration in a drone is the result of a complex interaction between mechanical, aerodynamic, and electronic forces. These vibrations are transmitted through the drone's structure (the chassis or frame) which, in some cases, not only propagates them but can even amplify them. To study this phenomenon accurately, four reflective markers were strategically placed on the top of the drone, as shown in Figure~\ref{fig:tackingUAV1}. These markers allow us to accurately track the dynamic behavior of the drone and analyze how vibrations affect its movement.

\subsection{Flight Characteristics}
\begin{figure}[ht]
    \centering
    \begin{subfigure}[b]{0.47\linewidth}
    \includegraphics[width=0.98\textwidth]{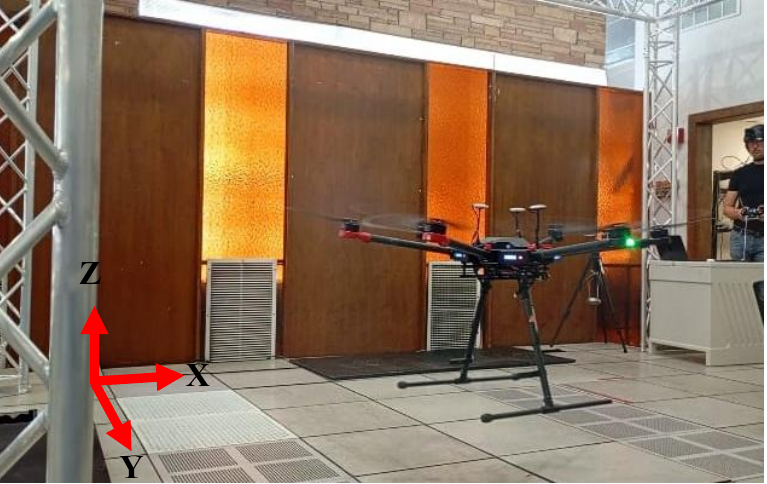}
    \subcaption{}\label{fig:charecteristicsb} \end{subfigure} 
    \begin{subfigure}[b]{0.49\linewidth}
    \includegraphics[width=0.98\textwidth]{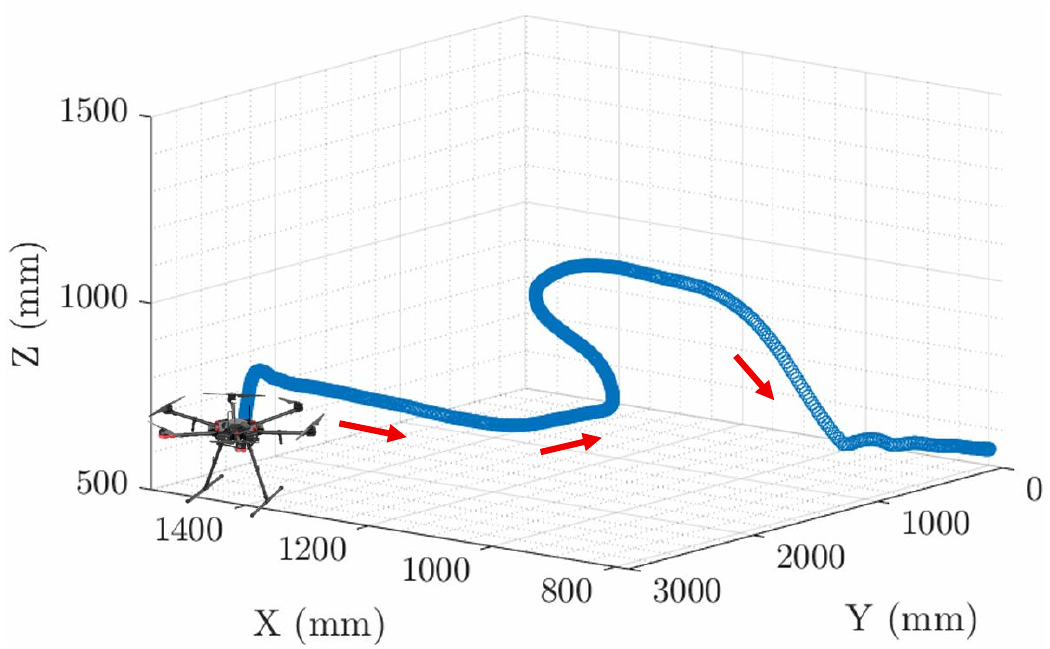}
    \subcaption{}\label{fig:fig:tackingUAVa} \end{subfigure}    
    \caption{Tracking UAV flight: (a) drone flight; (b) 3D view} \label{fig:trackingUAV_simulation}
\end{figure}

The data acquisition protocol was structured in three stages: initially, a calibration of the Vicon system was performed, verifying the accuracy of the optical tracking. The second stage involved the execution of a sequence of pre-designed movements, at different heights and with varied displacements to capture kinematic data during flight (see Figure~\ref{fig:charecteristicsb}). Finally, in the third stage, the dynamic characterization was carried out by reconstructing three-dimensional trajectories of the drone and calculating key kinematic variables such as displacements, velocities and accelerations, as you can see in Figure~\ref{fig:fig:tackingUAVa}. In this way, the dynamic behavior of the system under different operating conditions can be identified.

\subsection{Frequency Response} 

\begin{figure}[ht]
    \centering
    \begin{subfigure}[b]{0.35\linewidth}
    \includegraphics[width=0.98\textwidth]{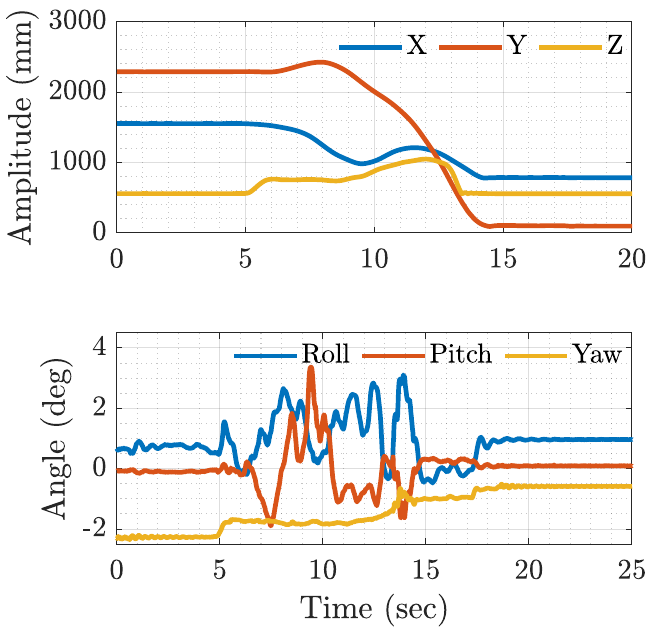}
    \subcaption{}\label{fig:fig:tackingUAVaa} \end{subfigure}
    \begin{subfigure}[b]{0.49\linewidth}
    \includegraphics[width=0.98\textwidth]{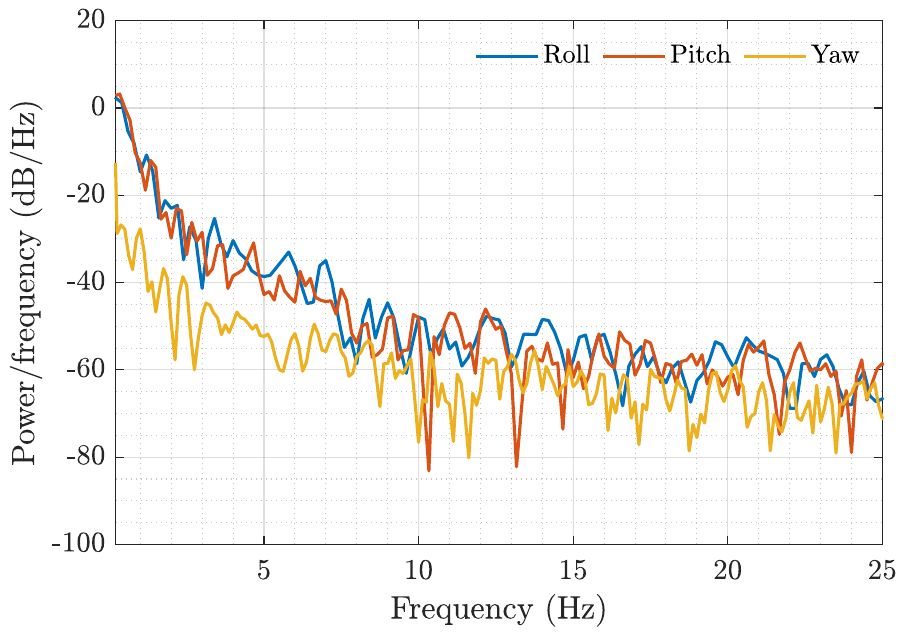}
    \subcaption{}\label{fig:fig:tackingUAVb} \end{subfigure} 
    \caption{Response a) Time domain ; b) Frequency response}\label{fig:freqDronea}
\end{figure}

Vibrations are an inherent phenomenon in the operation of drones and originate from multiple sources. As previously reviewed, the experiment was conducted in a laboratory environment; in this case, in addition to the vibrations generated mainly by the motors and propellers, the walls reflect the airflow produced, generating areas of turbulence and uneven pressure that increase aerodynamic drag.
In order to evaluate the behavior of the drone and detect natural frequencies, the experiment carried out with the Vicon cameras will record the displacements of the mobile platform in the three degrees of rotational and translational freedom. As the displacement in three dimensions has been recorded in Figure~\ref{fig:fig:tackingUAVaa}, the translation consists of moving from one end to the other, being the record in the time domain. 

When performing the analysis in the frequency domain, the natural frequencies obtained in Figure~\ref{fig:fig:tackingUAVb} are 0.5 Hz for roll, pitch, and yaw. According to observations under controlled conditions \cite{garg2020measuring}, the system's response to small disturbances shows that most of the flight power is concentrated in the frequency range below 0.5 Hz.

%% file: Sec4.tex

\section{Methodology}\label{method}


\subsection{Energy method}\label{energyMethod}

\begin{figure}[!ht]
    \centering
    \includegraphics[width=0.98\linewidth]{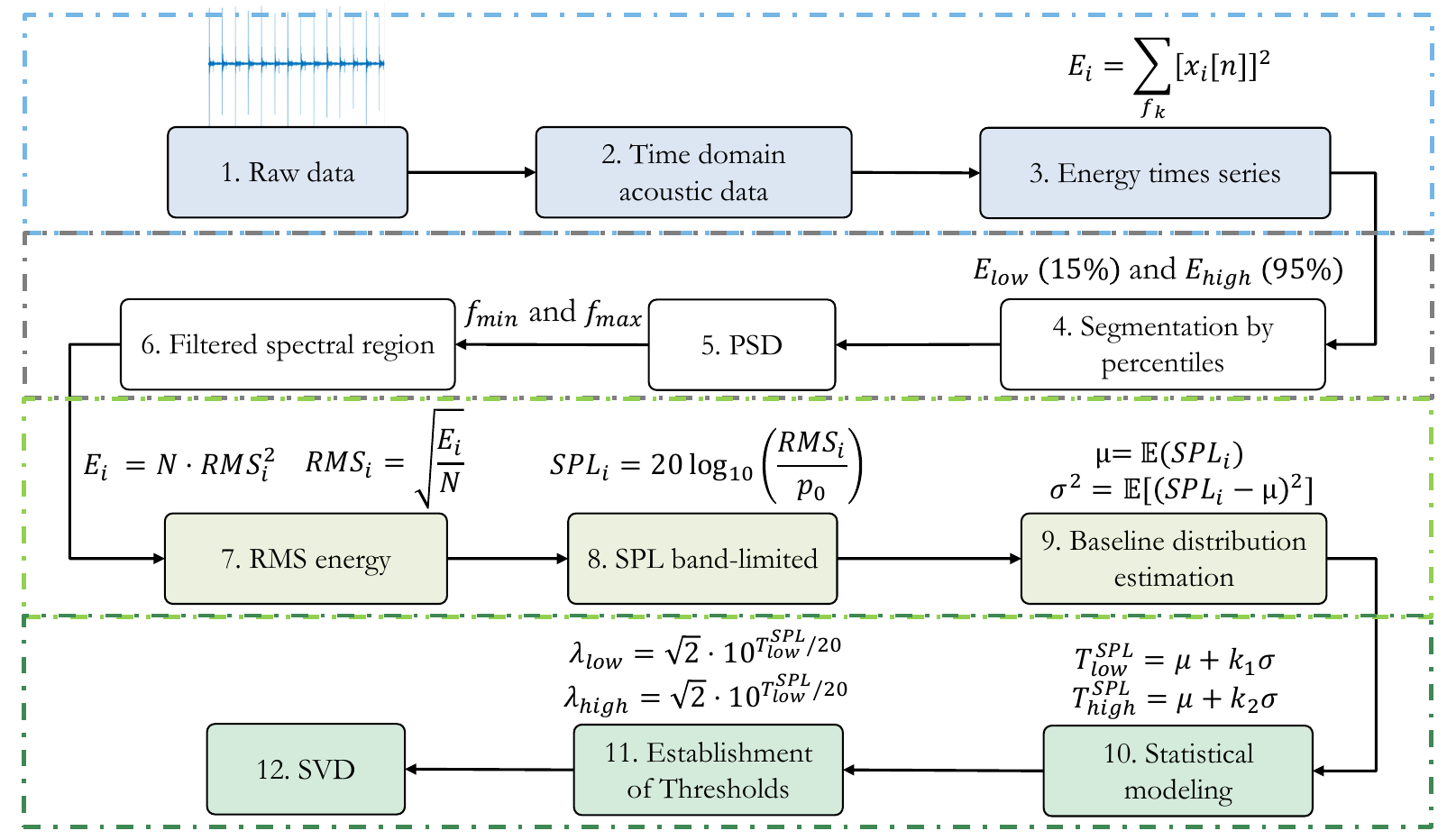}
    \caption{Energy method for defining minimum and maximum thresholds.}
    \label{fig:Methoddiagram}
\end{figure}

The analysis of acoustic time series in this study follows a signal processing workflow \cite{wang2025road}, the stages of which are described in Figure~\ref{fig:Methoddiagram}. 

The proposed procedure starts with the acquisition of the raw acoustic signal in the time domain, which constitutes the primary input for the analysis. The signal analysis is based on Parseval's theorem, which ensures that the energy calculated in the frequency domain is equivalent to that in the time domain \cite{brunton2022data}. The signal is segmented into short, overlapping time windows, and a short-time energy measure is computed for each window. This energy-based representation provides a descriptor of acoustic activity and allows the temporal evolution of the signal to be compactly characterized.
Percentile-based segmentation is applied to the energy distribution in order to distinguish between low-activity and high-activity regimes. Percentiles are not used as detection thresholds, but rather as a statistical tool to objectively separate background-dominated segments from event-dominated segments \cite{buxton2018efficacy}. This separation enables a data-driven spectral characterization without imposing arbitrary amplitude criteria on the original signal.

The PSD for each segment, \(P_{xx}^{(i)}(f)\), is calculated using the Welch method, which averages the spectra of multiple overlapping and windowed segments, providing a robust and smoothed estimate of the spectral distribution. 
Once the relevant spectral region has been identified, the energy within the band  of interest $([f_{\min}, f_{\max}])$ is computed for each segment as:
\begin{equation}
    E_i = \sum_{n=0}^{W-1} |x_i[n]|^2 = \sum_{k=0}^{K-1} |X_i[k]|^2 = \sum_{k=0}^{K-1} P_{xx}^{(i)}(f_k) \cdot \Delta f
\end{equation}
where \(\mathcal{B} = \{f_k : f_{\min} \leq f_k \leq f_{\max}\}\) is the selected frequency band, \(\Delta f\) is the frequency resolution, and \(f_k\) are the discrete frequency bins within \(\mathcal{B}\). 
The lower limit \(f_{\min}\) of the band was determined by identifying the frequency at which the spectral density decreases by 20 dB from the maximum observed in the low frequency range. This criterion automatically excludes the region where ambient noise dominates the spectrum, ensuring that only acoustically significant components are considered in the analysis. The upper limit \(f_{\max}\) is set where the spectral slope tends to zero, indicating the absence of significant spectral components.

Once the relevant spectral region has been identified, the acoustic signal is restricted to this band. A band-limited energy-consistent amplitude descriptor is then computed in the form of the root-mean-square (RMS) value over each time window. This RMS representation is subsequently converted into a band-limited Sound Pressure Level (SPL) time series, providing a logarithmic and physically meaningful measure of acoustic intensity.
From the band-limited SPL signal, a statistical baseline is established.
Based on this baseline statistical characterization, a statistical modeling stage is performed in which decision margins are defined as functions of the estimated mean and standard deviation.
Finally, lower (minimum) and upper (maximum) thresholds are established in the amplitude domain by mapping the statistically defined SPL thresholds back to the original signal scale. These thresholds provide a physically interpretable and statistically justified criterion for subsequent analysis and dimensionality reduction, including the application of singular value decomposition for feature extraction and pattern analysis.

\subsection{Post-processing} 

The acoustic data collected was analyzed using an algorithm that combines PCA, which provides a data-driven hierarchical coordinate system for representing high-dimensional correlated information \cite{brunton2022data} with a linear classification method, k-means, proposed by the authors of the study \cite{nasimi2021crack}.  
The selection of PCA and K-means was intentionally driven by the need for interpretability under controlled disturbance conditions. Unlike high-complexity models, these methods allow direct visualization of cluster structure and its evolution as vibration levels increase, enabling a clear assessment of how dynamic perturbations affect acoustic feature separability.


\input{Table/Algorithm_V2}

\subsubsection{Time domain procedure}
The analysis was performed on audio segments previously selected using the energy thresholding method, which correspond to high-energy events identified in the previous stage.


\subsubsection{Segmentation from acoustic data}
Next, relevant peaks are identified. The algorithm scans the subsampled signal $y$ to identify all local maxima (peaks) separated by at least $\delta$ units. These peaks are stored as pairs of amplitudes $v_k$ and positions $p_k$ in set $P$. Next, it filters $P$ by retaining only peaks whose amplitudes fall within the range $(\lambda_{min}, \lambda_{max})$, producing the final valid indices $P^*$. This ensures detected peaks meet both spacing and amplitude criteria. The output is a refined set of peaks ready for downstream analysis.
These segments are stored in a structured file, ready for further analysis. This systematic approach facilitating their use in classification or pattern recognition applications. 

\subsubsection{Data Split}
The training process starts with the split of the data, 60\% as a training set and 40\% as a testing set. A fundamental step in ML and related areas of data analysis is data partitioning. This stage usually consists of separating the data set into two distinct groups: training and testing. Its main purpose is to ensure accurate model evaluation and that the model can adequately generalize to new information. 


\subsubsection{PCA}
\begin{figure}[htbp]
    \centering
    \includegraphics[width=0.55\linewidth]{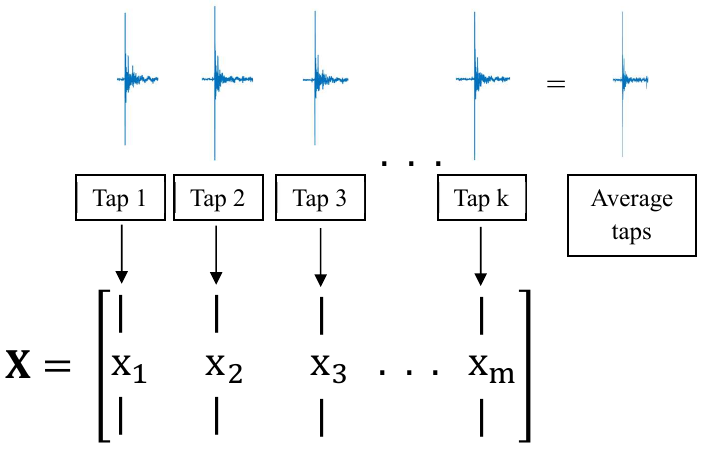}
    \caption{Mean-subtracted taps}
    \label{fig:MATRIXDAT}
\end{figure}

Next, a PCA is applied to the training data to reduce its dimensionality, retaining only those components that explain at least 90\% of the variance. This allows working with a more compact representation of the data without losing relevant information. PCA analysis uses the singular value decomposition (SVD) concept to find the principal components and the transformed data. Assuming that the A mxn is the matrix of data, where m represents the number of observations and n represents the number of the variables, the data matrix can be written as Equation~\ref{eq:PCA1}:

\begin{equation}\label{eq:PCA1}
    \text{\textbf{X}} = \text{\textbf{U}} \Sigma \text{\textbf{V}}^{T}  
\end{equation}
in which $\text{\textbf{U}}$ and $\text{\textbf{V}}^{T}$ are the left and right singular vectors, $\Sigma$ is a diagonal matrix which components’ is the square root of the Eigenvalues of $\text{\textbf{X}} \times \text{\textbf{X}}^{T}$ and 
the data matrix $\mathbf{X} \in \mathbb{X}^{n \times m}$, schematically represented in Figure~\ref{fig:MATRIXDAT}, contains in its columns the processed acoustic signals corresponding to each impact event recorded during the laboratory experiment. Specifically, the vector $\mathbf{x}_1$ contains the temporal response of the first hit, $\mathbf{x}_2$ captures the second impact, and this pattern extends sequentially up to the k-th event, $\mathbf{x}_m$, with respect to the rows, these contain the numerical information about each recorded peak. 

Using the singular value decomposition (SVD), the score matrix $\mathbf{T}$ can be written:

\begin{equation}\label{eq:PCA2E}
    \text{\textbf{T}} = \text{\textbf{X}} \text{\textbf{V}}^{T} = \text{\textbf{U}} \Sigma \text{\textbf{V}} \text{\textbf{V}}^{T}  =  \text{\textbf{U}} \Sigma
\end{equation}
so each column of $\text{\textbf{T}}$ is given by one of the left singular vectors of $\text{\textbf{X}}$ multiplied by the corresponding singular value. This form is also the polar decomposition of $\text{\textbf{T}}$.

\subsubsection{K-means}
Subsequently, K-means is used to group the transformed data into clusters. Each data point is then assigned to its nearest centroid based on Euclidean distance \cite{brunton2022data}, following the rule:

\begin{equation}\label{eq:kmean1}
    \mu_n = \text{argmin}_{\mu_j} \sum_{j=1}^k \sum_{h_i \in D_j'} \| h_i - \mu_j \|^2
\end{equation}

where $h$ is the i-th data point belonging to the current cluster and $\mu_j$ represents the centroid (or central position) of the j-th cluster during the clustering process. The centroids are subsequently recalculated as the mean of all points assigned to their cluster, mathematically expressed through the centroid update equation:

\begin{equation}\label{eq:kmean2}
    c_i = \dfrac{\Sigma^{N}_{n=1} h_n}{N} 
\end{equation}
where $N$ represents the count of points in the cluster. This process of point assignment and centroid recomputation continues iteratively until the cluster assignments stabilize, indicating convergence.

\subsubsection{Training ML Algorithms}

Finally, a decision tree classifier was selected to provide transparent and traceable decision boundaries, facilitating the interpretation of classification outcomes under varying vibration conditions.
In the test phase (40\% of the data), the PCA-transformed data are projected into the space. Next, the labels of the test set are estimated using the previously trained classifier and to evaluate the performance of the model, the predicted labels are compared with the actual labels by calculating metrics to quantify the performance of the trained model.

\subsection{Performance evaluation indexes} 

The following metrics (see Table~\ref{tab:Index}) are used as indicators in the evaluation of acoustic data classification performance \cite{goodfellow2016deep}. From the vibration data, the classification and its number of hits are extracted from the confusion matrix. The matrix is calculated for each case, where the true positive (TP), false positive (FP), false negative (FN) and true negative (TN) values are included. In this study, TP refers to the number of cases in which areas were correctly predicted to be in good condition; TN indicates the number of cases in which areas were correctly predicted to be peeling off the tile. In contrast, FP indicates the number of cases in which the areas that were in good condition were predicted incorrectly; FN refers to the number of cases in which the areas that were peeling off were predicted incorrectly.

\input{Table/Metrics}

%% file: Table/Algorithm_V2.tex
\begin{algorithm}[th!] 
\footnotesize
\caption{Sequence of the signal processing and classification algorithm }\label{alg:PCAkmeans}

\begin{algorithmic}[1]
 \renewcommand{\algorithmicrequire}{\textbf{Input:}}
 \renewcommand{\algorithmicensure}{\textbf{Output:}}
 \REQUIRE $x$ audio signals and $F_s$ sampling frequency
 \ENSURE  \(\text{mdl}\) Trained decision tree model and Cluster assignments 

\STATE  Extract sound data \\
    For each signal \( x \): Extract segments \( y = \{ x_1, x_1, . . ., x_n \} \)  

\STATE  Data Segmentation \\

    For each subsampled identifies peaks in signal $y$ with minimum spacing $\delta$: \\ 
    \(P = \{ (v_k,p_k) | v_k = y(p_k), p_k \in \text{MaxLocal}(y), |p_k-p_{k-1}| \geq \delta \} \) \\
    Keeps peaks within amplitude bounds \( \lambda_{min}, \lambda_{max} \):  \\
    \(P^* = \{ p_k|(v_k,p_k) \in P, \lambda_{min} < v_k < \lambda_{max} \} \) \\

    Segments of fixed length \(L\) are extracted are extracted along the amplitude \(v_k\) of each detected peak. Combine all segments (peaks) into a numerical matrix \( X = \text{stack}(x_1,x_2,...,x_n) \), where \(x_i\) is each segment. 
    
\STATE Data preparation\\
    Divide data \(X\) into 60\% training and 40\% testing. \\
    
\STATE  PCA \\
    Compute mean row \(\bar{X}\) and subtract mean \(\bar{X}\) from \(X\) data, \(B = X - \bar{X} \) \\
    Covariance matrix of rows of \(B\), \(C = B^{T}B\) \\ 
    And then obtain the principal components by computing the eigen decomposition of C, where \(CV = VD\) contains the eigenvectors called the loadings \(V\) and their eigenvectors \(D\).
    
\STATE  Clustering with K-means \\
    Initialize centroid position and assign labels \(\mu\) to all data \(H\), like Eq.~\ref{eq:kmean1} \\
    Update centroid postions \(C\) with Eq.~\ref{eq:kmean2} and repeat until convergence 

\STATE Classifier training\\
    Train decision tree \(\text{mdl}\)

\STATE \textbf{Return:} Trained neural network

\end{algorithmic} 
\end{algorithm}

%% file: Table/Metrics.tex
\begin{table}[!ht]
    \centering
    \footnotesize
    \caption{Evaluation metrics of the classification model}
    \begin{tabular}{|m{0.30\linewidth} | m{0.6\linewidth}|}
    \hline
    Description & Equation \\ 
        \hline
        \begin{equation}  \frac{TP}{TP+FP}\end{equation}\label{eq:precision} &     \textbf{Precision}: Proportion of correct positive predictions out of all positive predictions. 
        \\         
        \hline
        \begin{equation}  \frac{TN}{TN+FN} \end{equation}\label{eq:NVP} & \textbf{Negative Predictive Value (NPV)}: Proportion of correct negative predictions over all negative predictions. 
        \\
        \hline
        \begin{equation} \frac{TP}{TP+FN} \end{equation}\label{eq:NVP} & \textbf{Sensitivity (Recall)}: Proportion of correctly identified positive cases among all true positives. 
        \\ 
        \hline
        \begin{equation}  \frac{TN}{TN+FP} \end{equation}\label{eq:NVP} & \textbf{Specificity}: Proportion of correctly identified negative cases among all actual negative cases. 
        \\ 
        \hline 
        \begin{equation} \frac{TP + TN}{TP+TN+FP+FN} \end{equation}\label{eq:NVP} & \textbf{Accuracy}: Proportion of correct predictions (positive and negative) over all predictions made. 
        \\ 
        \hline
    \end{tabular}

    \label{tab:Index}
\end{table}

%% file: Sec3.tex

\section{Experiments integration}

\subsection{Stewart Platform} 

The Stewart platform is a versatile parallel robot interconnected by six independently controlled linear actuators \cite{silva2022stewart}. Its design allows for movements in 6 DOF, making it ideal for simulation and testing applications. In this study, the Stewart platform was used to simulate controlled vibrations in order to reproduce disturbances characterized by varying amplitudes and frequencies. This capability is especially suited to the experiment to evaluate the effectiveness of tap tests, a non-destructive testing method commonly used to assess the structural integrity of materials and components.

\begin{figure}[ht] 
    \centering
    \includegraphics[width=0.85\linewidth]{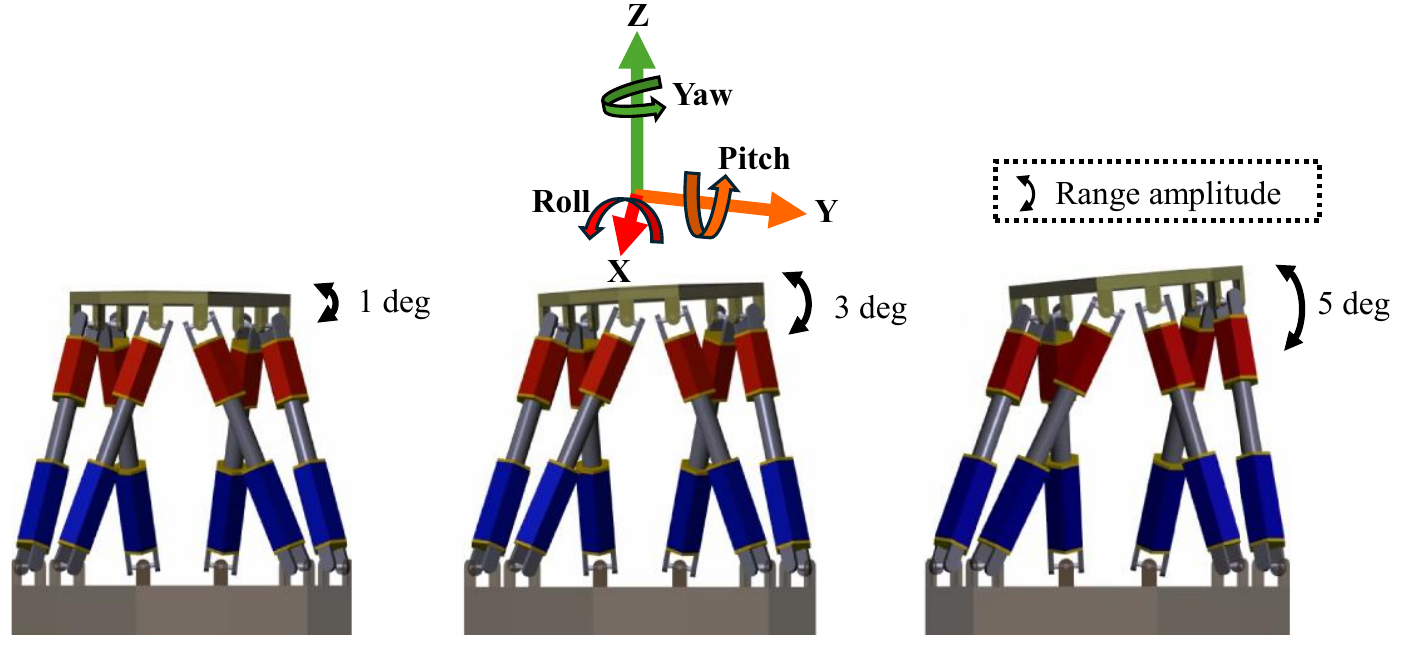}
    \caption{Experiment with vibrations}
    \label{fig:hexapod3levels}
\end{figure}

The controlled oscillations were specifically designed to reproduce the turbulence or disturbances that a drone might encounter during flight, allowing for a realistic simulation of the vibrations experienced by the UAV in various flight scenarios \cite{vibrationHexapod}. In this experiment, as illustrated in Figure~\ref{fig:hexapod3levels}, three levels of amplitude were established, keeping the frequency constant. The only variable between the cases was the amplitude of the wave, which was then compared with the base experiment (tap testing without movement).

\begin{figure}[!ht] 
    \centering
    \includegraphics[width=0.85\linewidth]{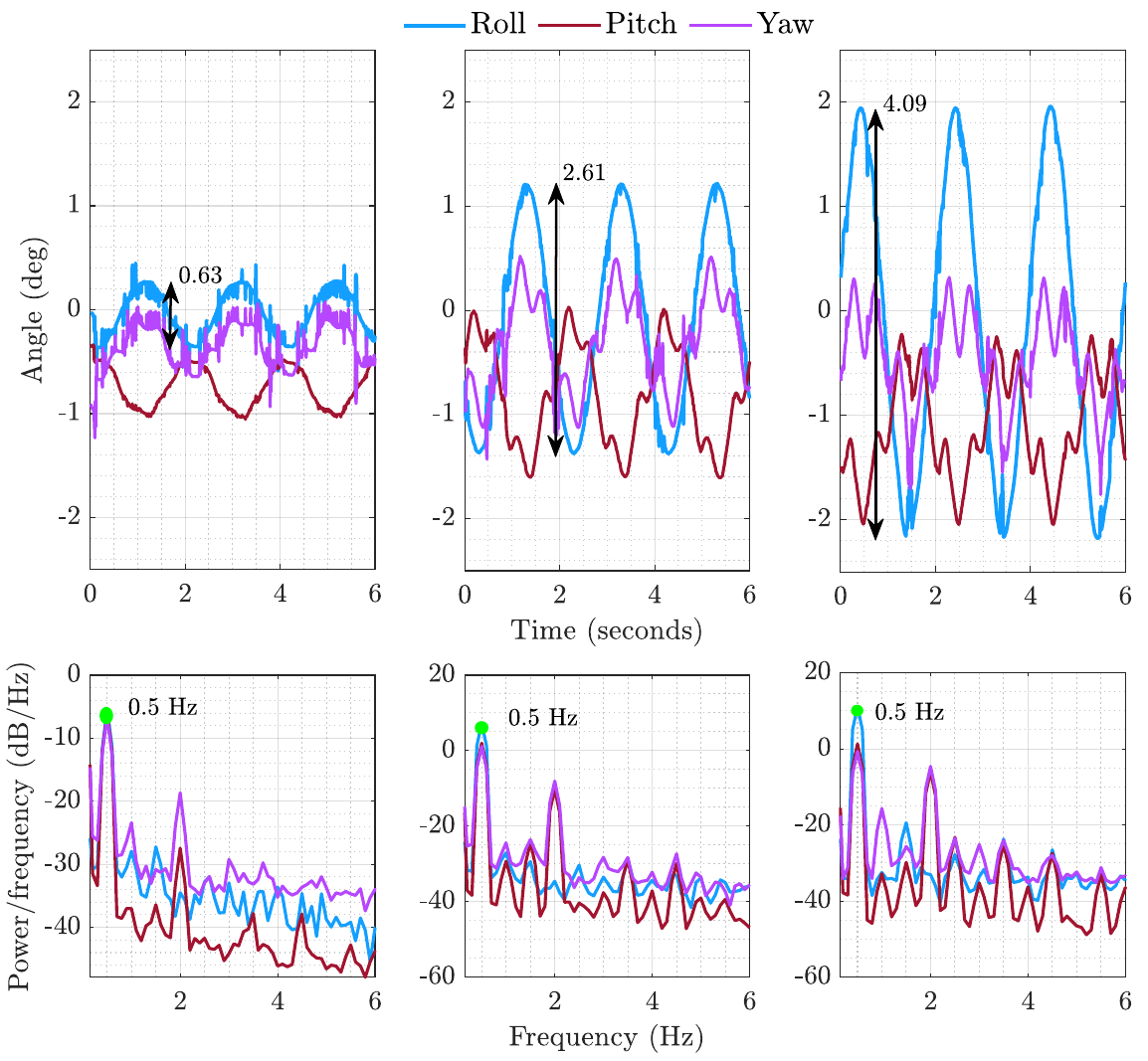}
    \caption{Time and frequency domain}
    \label{fig:hexapodFreq}
\end{figure}

For each angular time series (Roll, Pitch, and Yaw), the curves in In Figure~\ref{fig:hexapodFreq} are analyzed to locate maximum points (peaks) and minimum points (valleys). These maximum and minimum points represent the amplitude range at which they were established, that is, an approximation of the actual value of 1, 3, and 5 degrees. 
The first plot shows the three curves with sinusoidal behavior and slight noise, with amplitudes close to 1 degree. In the second and third plots, only roll maintains a sinusoidal shape, while pitch and yaw show variations in their behavior.

Table~\ref{table: degMov} provides the details of the degrees of vibration applied during the simulation, along with their corresponding periods, over a duration of 90 seconds. The results of the three experiments with different amplitudes, as mentioned previously, have the same input frequency, and the output recorded in all three cases is 0.5 Hz. 

\begin{table}[!ht]
    \centering \footnotesize
    \caption{Frequencies identified in each case experiment using tracking cameras}
    \begin{tabular}{|m{2.1cm}|m{1.6cm}|m{1.8cm}|m{2.2cm}|m{2.2cm}|m{2.2cm}|} \hline
     & \multicolumn{5}{c|}{\textbf{Frequency (Hz)}} \\ \hline
    \textbf{Rotational DOF} & \textbf{Identified} & \textbf{Hexapod input} & \textbf{1 deg output} & \textbf{3 deg output} & \textbf{5 deg output} \\  \hline
    Roll & 0.5 & 0.5 & 0.5 & 0.5 & 0.5 \\ \hline
    Pitch & 0.5 & 0.5 & 0.5 & 0.5 & 0.5 \\ \hline
    Yaw & 0.5 & 0.5 & 0.5 & 0.5 & 0.5 \\ \hline 
    \end{tabular} \label{table: degMov}
\end{table}


\subsection{Specimen}

In order to evaluate the structural and adhesion behavior of wall tiles under varying conditions, a series of controlled laboratory tests were conducted using representative specimens designed to replicate real-world building facade environments. 
The specimen's dimensions were carefully selected to ensure geometric and mechanical similitude with actual construction conditions. 

\begin{figure}[ht]
    \centering
    \includegraphics[width=0.7\linewidth]{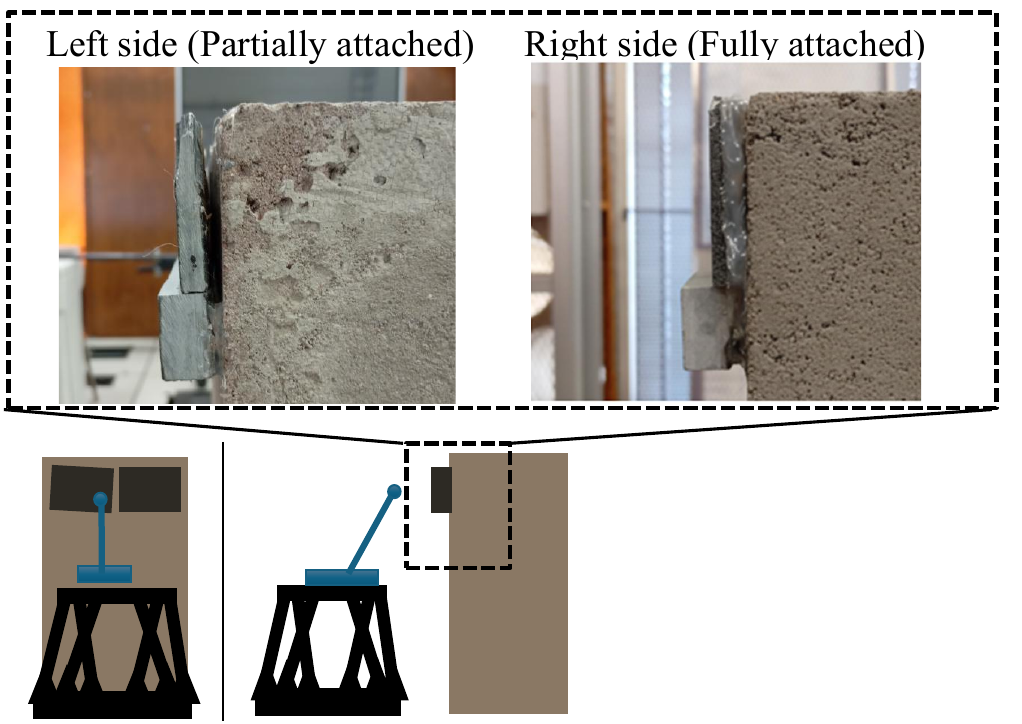}
    \caption{Specimen Elevation View}
    \label{fig:specimenElevation}
\end{figure}

The experimental setup involved mounting the tiles onto standardized blocks, with a deliberate debonding defect introduced by applying adhesive to only 50\% of the tile's surface area.

\begin{table}[htbp]
    \centering
    \footnotesize
    \caption{Properties of material}
    \begin{tabular}{|l|m{4cm}|m{5.3cm}|} \hline
    \textbf{Materials} & \textbf{Size} & \textbf{Characteristic} \\ \hline
        Tile & 8 $\times$ 3 $\times$ 1/4 in$^{3}$ & Charcoal Ledgestone stone \newline Textured/rough \newline High wear resistance\\
        \hline
        Adhesive layer & 5 $\pm$ 0.5 mm & Structural silicone and epoxy resin adhesive coating \\  \hline
        Block & 15 $\times$ 7 1/2 $\times$ 7 1/2 in$^{3}$ & High density precast concrete block \\
        \hline
    \end{tabular}
    \label{tabC:tile}
\end{table}

This partial bonding methodology was implemented to simulate common failure modes observed in deteriorated facades, thereby allowing for systematic analysis of detachment mechanisms. The materials employed in this investigation, including tile composition, adhesive properties, and substrate characteristics, are comprehensively documented in Table~\ref{tabC:tile} to ensure reproducibility and facilitate comparative analysis with existing studies.

\subsection{Setup}

A remote-controlled robot for the automation of acoustic data collection in potential rockfall studies was developed and validated in the field \cite{nasimi2022use}. In this work, the tap testing system is transferred to a controlled laboratory environment, where a 6 DOF platform is integrated as a key component to evaluate the system performance under simulated dynamic perturbations. 

\begin{figure}[htbp]
    \centering \includegraphics[width=0.75\linewidth]{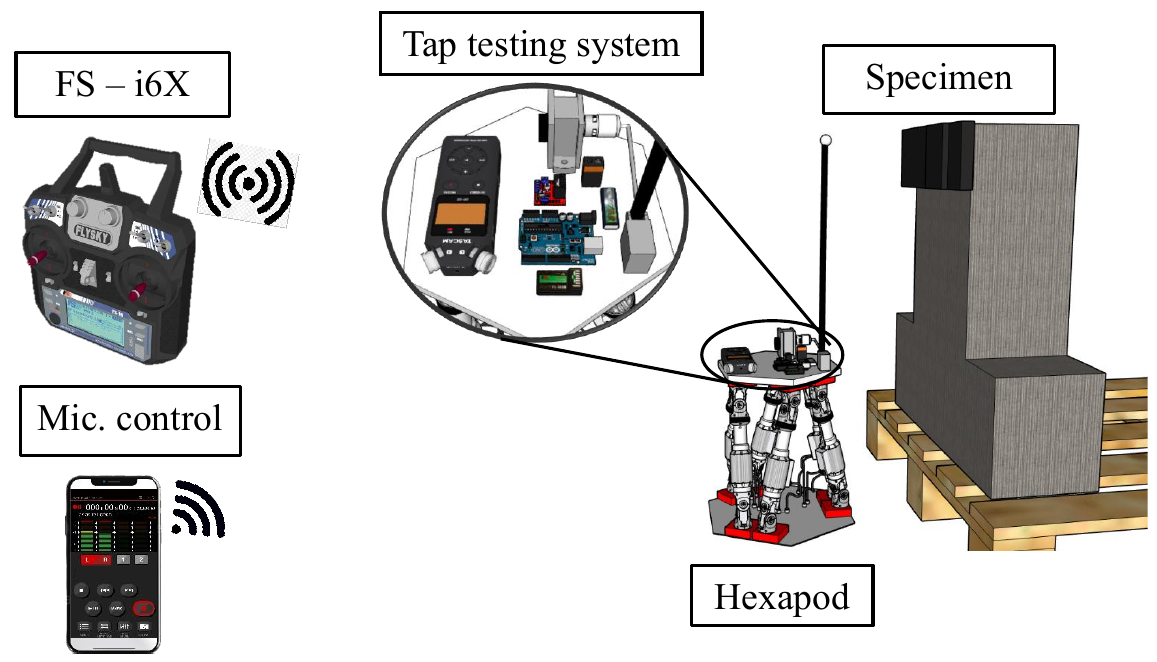}
    \caption{Experimental setup}
    \label{fig:Setup}
\end{figure}

The system consists of a tapping mechanism that generates a controlled force on the specimen surface, producing a characteristic acoustic response. This signal is captured by a highly sensitive microphone, whose data stream is transmitted remotely via WiFi through an interface controlled from a mobile device. 

This experimental phase lies in the incorporation of the 6DOF platform, which introduces vibrations to emulate realistic dynamic conditions, such as vibrations during drone flight. Figure~\ref{fig:Setup} illustrates the laboratory experimental setup, where the platform allows simulating controlled perturbations in the data acquisition system, replicating different levels of operational uncertainty. Subsequently, the information will be evaluated with the prediction algorithm against variations in the amplitude, frequency and direction of the induced vibrations. Table~\ref{tab1 : BRUTUS_system} summarizes the main elements that make up the conceptual illustration.

\begin{table}[!hbt]
    \centering
    \footnotesize
    \caption{Components of the experimental setup}
    \begin{tabular}{|m{3.5cm}|m{3.5cm}|m{7cm}|}
        \hline
    \textbf{Component} & \textbf{Specifications} & \textbf{Purpose}\\
        \hline
        Main board & Arduino UNO R3 & An input–output device used to help control other components of BRUTUS by utilizing personally developed code.\\
        \hline
        Motor & TSINY & Drives the rocker arm mechanism. \\
        \hline
        Mechanism & Crank-Rocker & Drives tap testing hammer head; hammer head creates acoustic response (Steel Ball Knob). \\
        \hline
        Driver Motor Controller & L298N 2A & Gives full control of motor speed and function through communication via the Arduino Uno. \\
        \hline
        Transmitter & FS-i6X Transmitter & Used for remote control of the BRUTUS device; communicates with device through the FS-iA6B receiver. \\
        \hline
        Receiver & FS-iA6B receiver &  Relays information between FS-iA6B transmitter \\
        \hline
        Batteries & 9 V and 7.4 V & Used to power BRUTUS. \\
        \hline
        Microphone & TASCAM PCM recorder & Record and stores acoustic response data. \\
        \hline
        Hexapod & 6 DOF & Provide movement to the system. \\
        \hline
    \end{tabular}
    \label{tab1 : BRUTUS_system}
\end{table}

\subsection{Experimental procedure}

The experimental setup was designed to capture controlled acoustic sounds from a composite specimen under mechanical excitation. The specimen was mounted: the left side was partially attached to simulate a loosened or degraded connection, while the right side was fully attached representing an ideal structural joint. Mechanical impacts were delivered using a tap testing system synchronized with a hexapod platform, which provided reproducible multi-axis excitation. Acoustic responses were recorded with a condenser microphone, positioned 30 cm from the specimen to optimize signal capture. The microphone signal was conditioned through a dedicated microphone control unit and digitized at 44.1kHz to ensure wideband acquisition. This configuration allowed the study of acoustic emission patterns under varying stiffness conditions, with precise synchronization between impact triggers and acoustic data collection.

%% file: Sec6.tex

\section{Results}

\subsection{Zero Vibration condition} \label{zerocondS}
\begin{figure}[!htbp]
    \centering
    \begin{subfigure}[b]{0.42\linewidth}
    \includegraphics[width=0.98\textwidth]{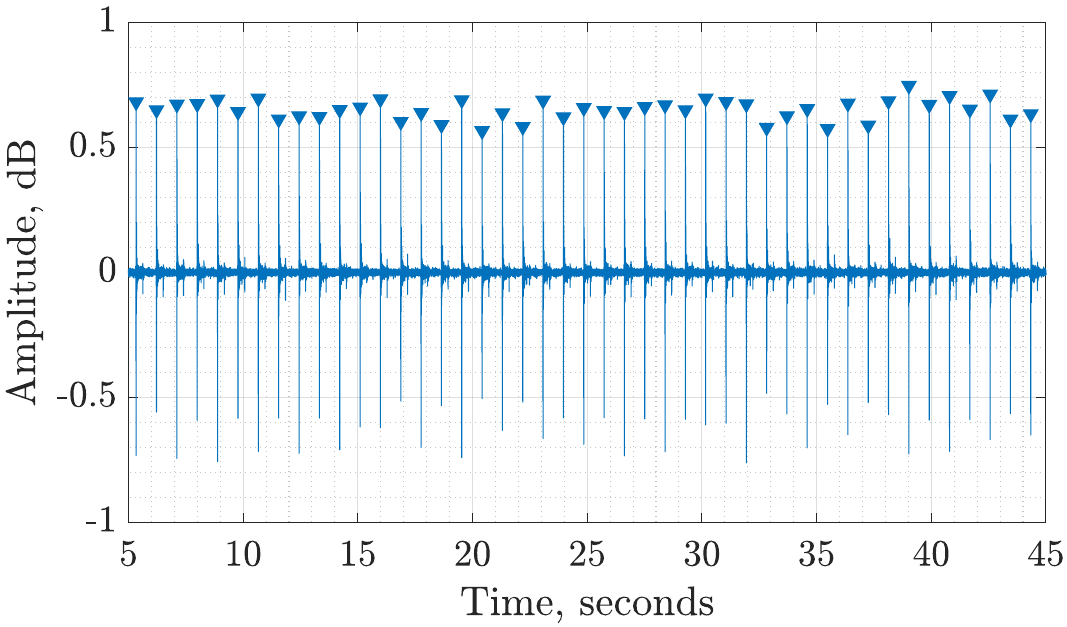}
    \label{fig:a} \subcaption{}
    \end{subfigure}
    \begin{subfigure}[b]{0.42\linewidth}
    \includegraphics[width=0.98\textwidth]{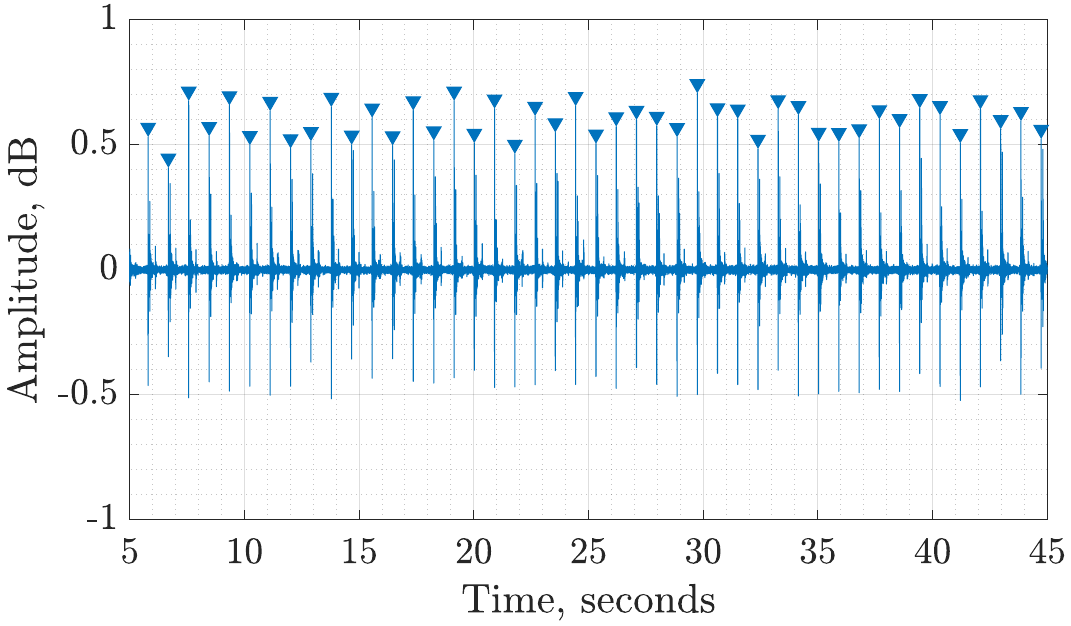}
    \label{fig:b} \subcaption{}
    \end{subfigure} 
    \caption{Acoustic data: (a) Healthy condition; (b) Unhealthy condition} \label{fig:timehistory}
\end{figure}

The acoustic data recorded in the laboratory without movement were very clear, where an amplitude difference is visualized. First, to evaluate the recorded sound data, time histories were plotted as shown in Figure~\ref{fig:timehistory} for both study cases (healthy and unhealthy condition)

To evaluate the performance of the algorithm in identifying adhered and partially adhered tiles, a stratified division of the data set was implemented, assigning 60\% of the samples of each type of surface for model training and reserving the remaining 40\% for the validation phase, as detailed in Table \ref{tabC:division}.

\begin{table}[!htbp]
    \centering
    \footnotesize
    \caption{Information of the training and testing data.}
    \begin{tabular}{|l|c|c|c|} \hline
    Experiment & Training taps & Testing taps & Correctly classified  \\ \hline
        Good condition & 64 & 38 & 38 \\ \hline
        Bad condition & 64 & 39 & 39 \\ \hline
        Total taps & 128 & 77 & 77 \\
        \hline
    \end{tabular}
    \label{tabC:division}
\end{table}

\begin{figure}[!ht]
    \centering
    \begin{subfigure}[b]{0.4\linewidth}
    \includegraphics[width=0.98\textwidth]{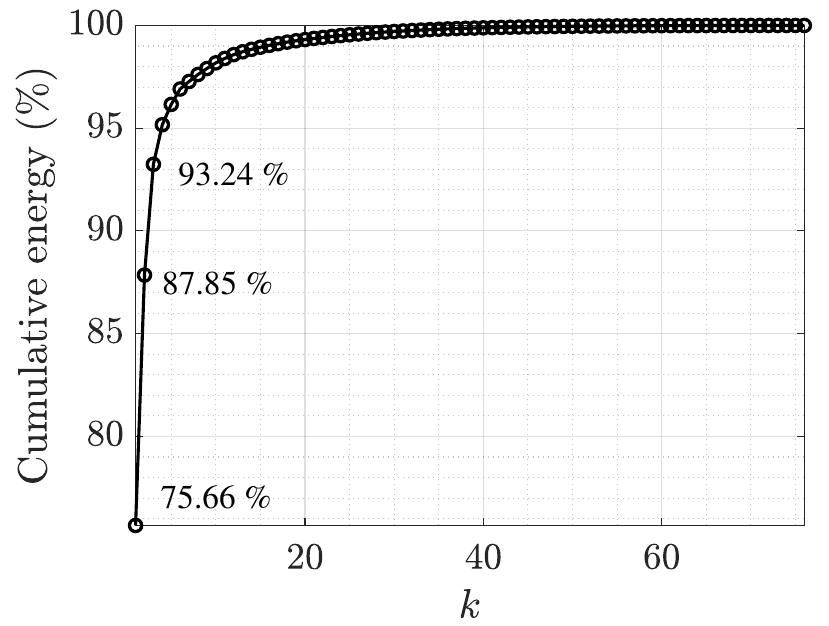}
    \subcaption{}\label{fig:PCA_a}
    \end{subfigure}
    \begin{subfigure}[b]{0.41\linewidth}
    \includegraphics[width=0.98\textwidth]{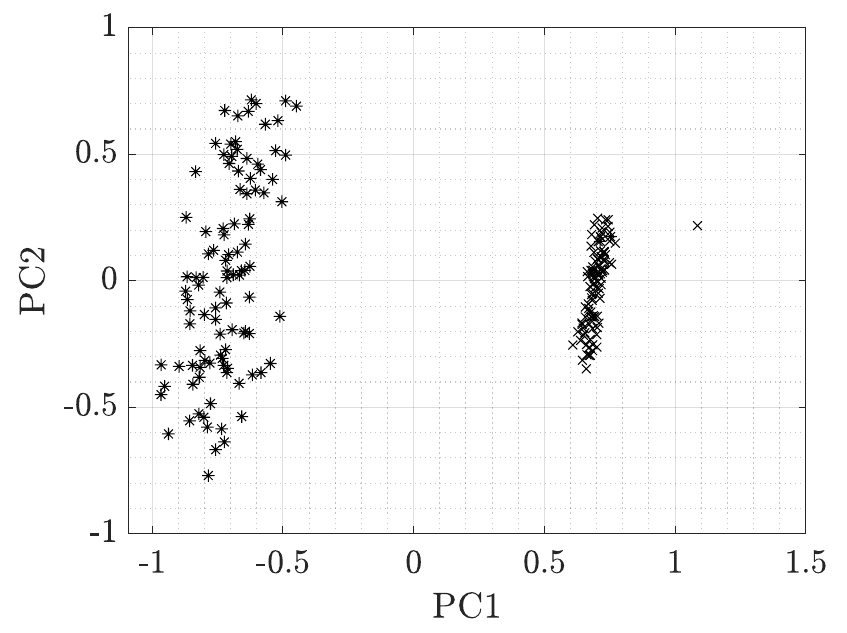}
    \subcaption{}\label{fig:bPCA} 
    \end{subfigure} 
    \caption{Acoustic data with PCA: (a) Cumulative energy in first $k$ modes; (b) Clustering of samples that are normal and those that are unhealthy condition in the first two principal component coordinates.} \label{fig:energyPCA}
\end{figure}

For each experimental set composed of 180 acoustic impacts, it is necessary to reduce the dimensionality before proceeding to the classification of the tile impacts. PCA is applied, paying special attention to the explained variance of the first modes \cite{abdi2010principal}. The Figure~\ref{fig:PCA_a} shows the cumulative energy (or variance explained) as a function of the number of principal components k. As can be seen, the first two modes manage to capture a significant variance of 87\% of the total data. In other words, Figure~\ref{fig:bPCA} shows that the impacts on unhealthy and healthy tiles are clustered on opposite sides of the PC1 axis (left and right, respectively). The difference between their means in PC1 is about 1.4 units, indicating a significant linear separation between the classes. In PC2 the distribution is less differentiated, the two groups overlap more, but may provide secondary information.

\begin{figure}[!htbp]
\centering
\begin{subfigure}[b]{0.4\linewidth}
\includegraphics[width=0.98\textwidth]{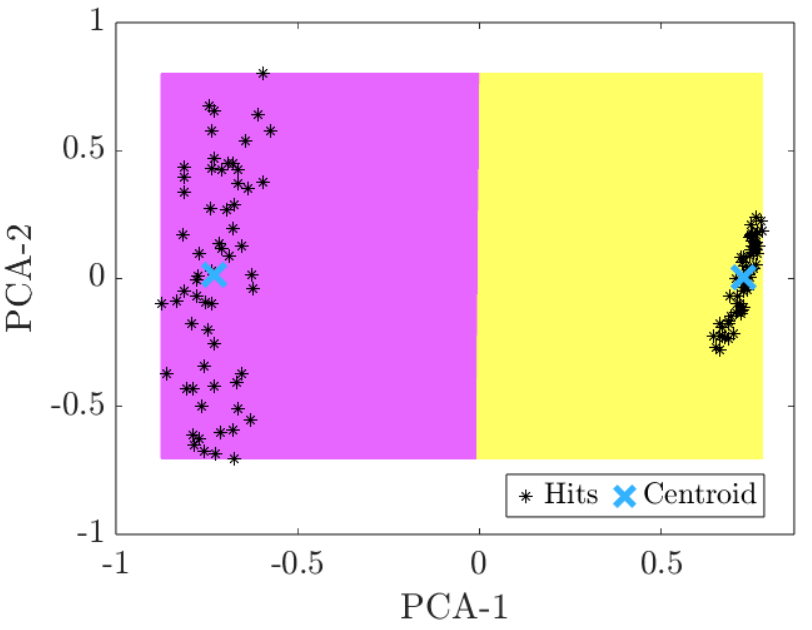}
\subcaption{}\label{fig:trainingD}
\end{subfigure}
\begin{subfigure}[b]{0.4\linewidth}
\includegraphics[width=0.98\textwidth]{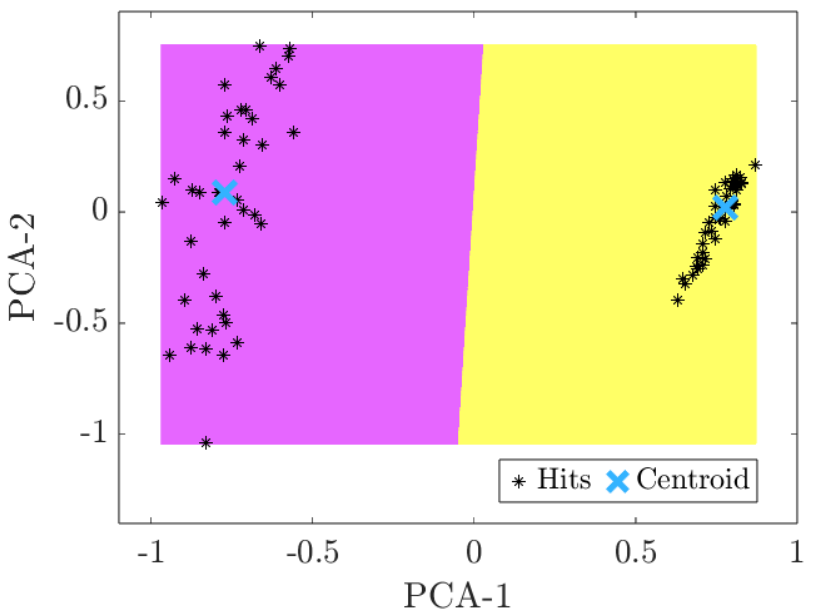}
\subcaption{}\label{fig:testingD}
\end{subfigure} 
\caption{Data classification: (a) Training data; (b) Testing data}
\label{fig:entrenamientopruebas}
\end{figure}

The cluster centers were then determined using the k-means algorithm, the training data were classified and the characteristic regions were defined, as shown in Figure~\ref{fig:trainingD}.
The previously implemented model is evaluated with the test set to assess its classification capability, the results of which are presented in Figure~\ref{fig:testingD}.

\subsection{Vibration condition}

\begin{figure}[!h]
\centering 
\begin{subfigure}[b]{0.33\linewidth}
\includegraphics[width=0.99\textwidth]{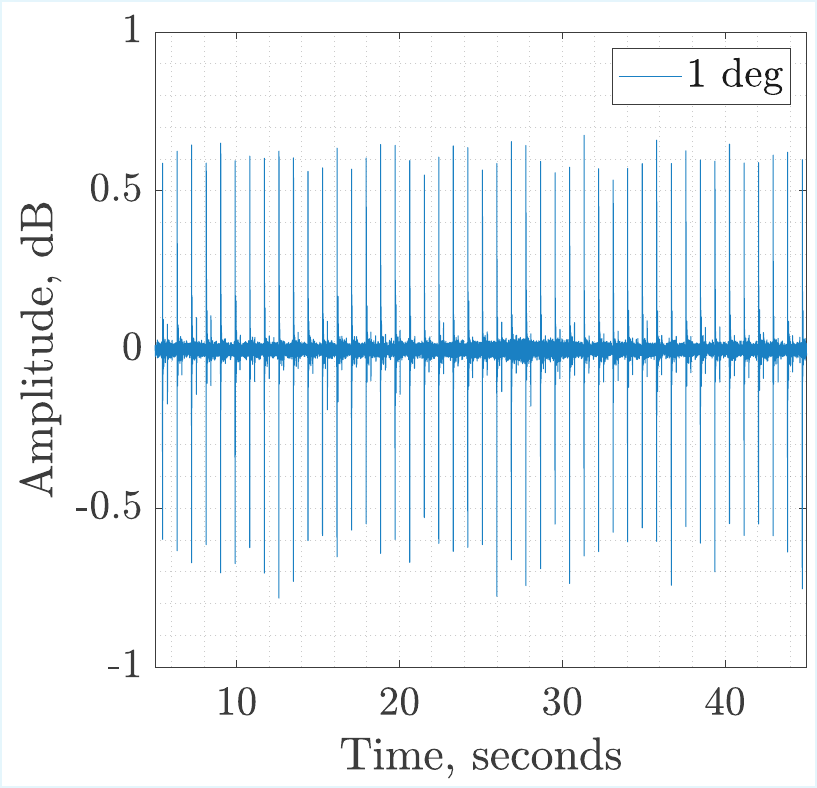}
\label{fig:a} \subcaption{} \end{subfigure}
\begin{subfigure}[b]{0.31\linewidth}
\includegraphics[width=0.99\textwidth]{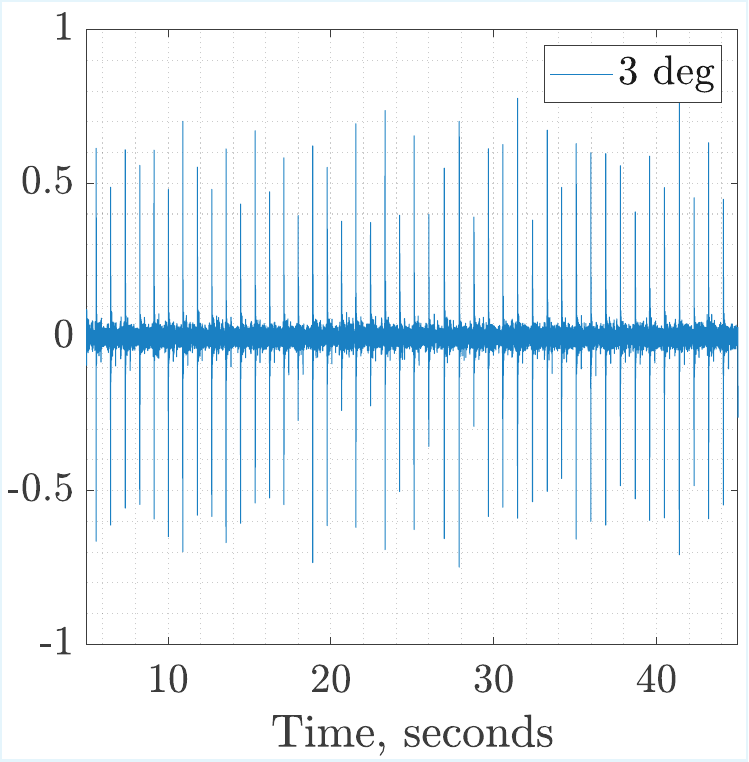}
\label{fig:b} \subcaption{} \end{subfigure} \hspace{-3mm}
\begin{subfigure}[b]{0.30\linewidth}
\includegraphics[width=0.99\textwidth]{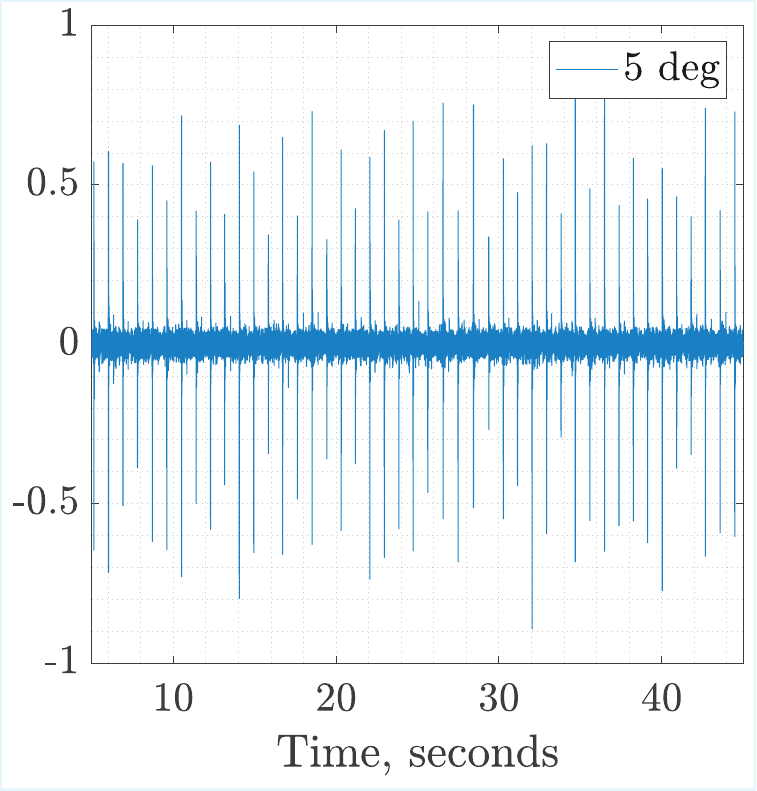}
\label{fig:b} \subcaption{} \end{subfigure}
\caption{Time domain healthy conditions (a) 1 deg; (b) 3 deg; (c) 5 deg} \label{fig:resultTotalHcBad}
\end{figure}

The study continues with the induction of controlled vibrations, those three levels mentioned above of 1, 3, and 5 degrees. The three sets of experiments that increase vibration can be visualized in the time domain in Figure~\ref{fig:resultTotalHcBad}.

\begin{figure}[htbp]
\centering 
\begin{subfigure}[b]{0.33\linewidth}
\includegraphics[width=0.99\textwidth]{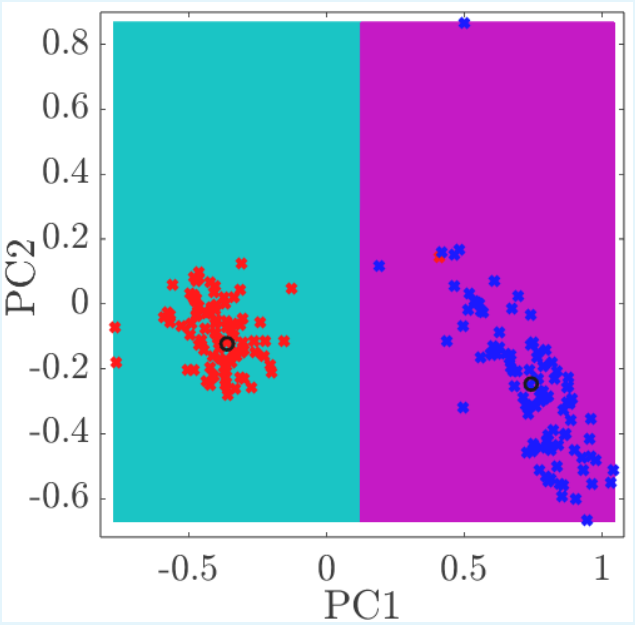}
\label{fig:a} \subcaption{} \end{subfigure}
\begin{subfigure}[b]{0.295\linewidth}
\includegraphics[width=0.99\textwidth]{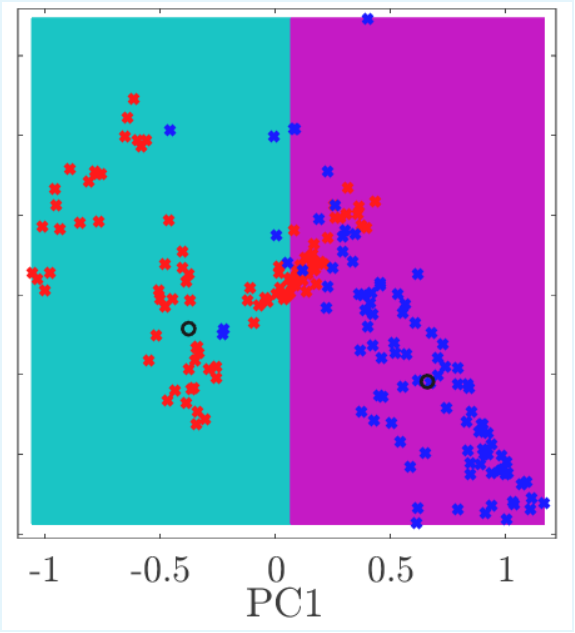}
\label{fig:b} \subcaption{} \end{subfigure} \hspace{-3mm}
\begin{subfigure}[b]{0.33\linewidth}
\includegraphics[width=0.99\textwidth]{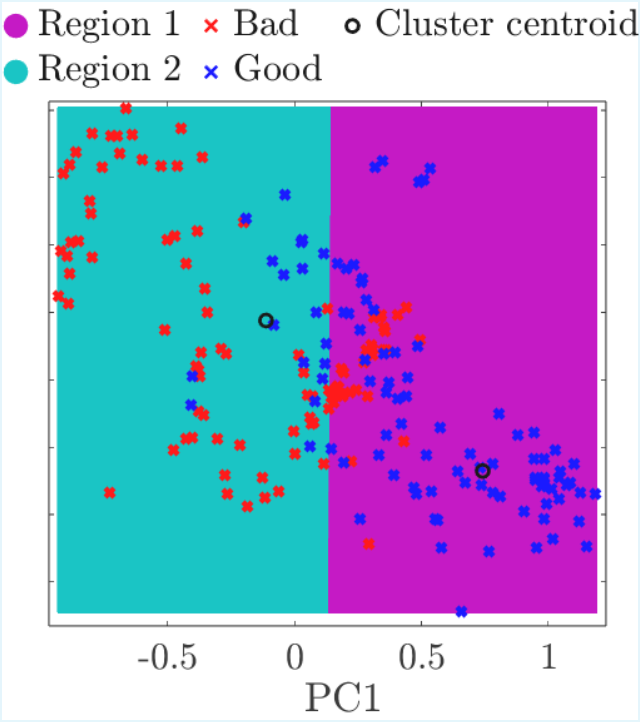}
\label{fig:b} \subcaption{} \end{subfigure}
\caption{PCA (a) 1 deg; (b) 3 deg; (c) 5 deg} \label{fig:resultTotalPCA}
\end{figure}

Under controlled conditions, the vibrational inputs introduce measurable perturbations to the acquired acoustic data, as evidenced by the clustering patterns in the Figure~\ref{fig:resultTotalPCA}. The PCA reveals varied groupings corresponding to specific vibrational modes indicating less compact clusters, with a clear dispersion from the baseline (no motion) results. This suggests that external vibrations systematically alter the signal structure, significantly compromising the accuracy of damage-related acoustic feature detection.

Table~\ref{tabC:divisionwithVibrationv1} suggests that the model maintains acceptable specificity but loses sensitivity as the vibration intensity increases. The progressive decrease in hits  and increase in errors indicate that the variability generated by more intense vibrations significantly affects the model's ability to correctly identify the faulty condition.

\begin{table}[!htbp]
    \centering \footnotesize 
    \caption{Results of the performance evaluation for training a testing.}
    \begin{tabular}{|c|c|c|c|c|c|c|c|c|}
    \hline
                   & \multicolumn{2}{c|}{0 deg} & \multicolumn{2}{c|}{1 deg} & \multicolumn{2}{c|}{3 deg} & \multicolumn{2}{c|}{5 deg} \\  \hline 
                   Condition & Good & Bad & Good & Bad & Good & Bad & Good & Bad \\  \hline
    Total Taps  & 38 & 39 & 96 & 97 & 96 & 97 & 96 & 97 \\ \hline
    Correctly classified   & 38 & 39 & 96 & 96 & 92 & 57 & 90 & 50 \\ \hline
    Incorrectly classified & 0 & 0 & 0 & 1 & 5 & 39 & 6 & 47 \\ \hline
    \end{tabular}
    \label{tabC:divisionwithVibrationv1}
\end{table}

This highlights the need to increase the diversity of the training data set, for cases with medium and high vibrations, and to consider regularization techniques or more complex architectures that can better handle the variability introduced by different levels of perturbation.

\subsection{Comparison of vibration level performance}

\begin{figure}[htbp]
    \centering
    \includegraphics[width=0.55\linewidth]{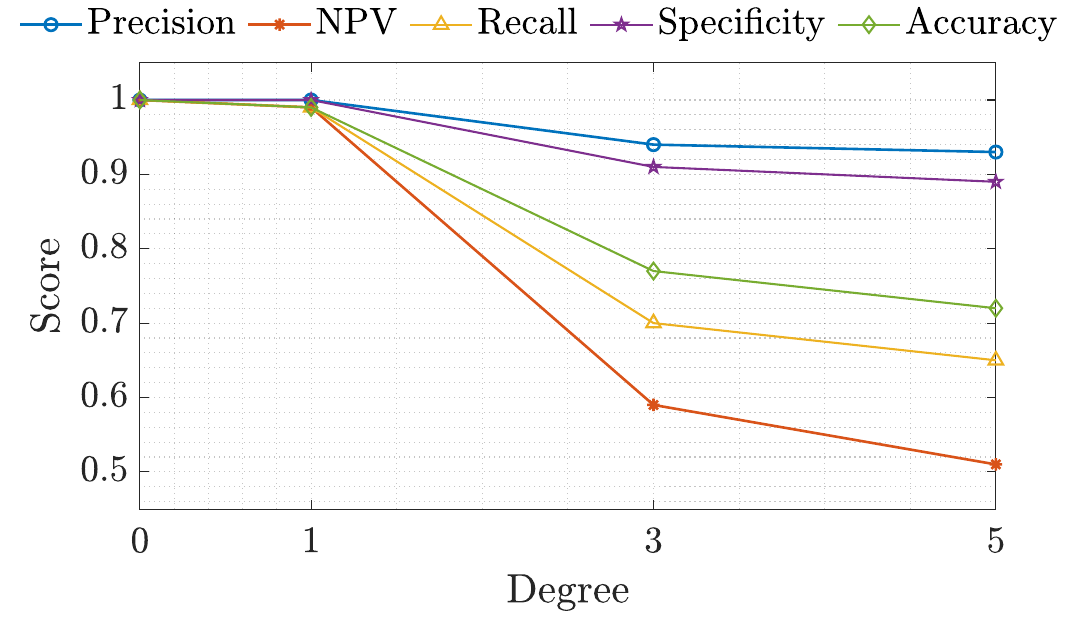}    
    \caption{Base performance and three vibration levels}
    \label{fig:generalScoreBefore}
\end{figure}

In order to establish a qualitative comparison, the evaluation process of the previously established conditions is presented below. 
The performance can be seen visually in Figure~\ref{fig:generalScoreBefore}, the same as that shown in Table~\ref{tab:generalScore1deg} with detailed rating values.

\input{Table/tableGeneralResult}

\begin{figure}[htbp]
    \centering
    \includegraphics[width=0.30\linewidth]{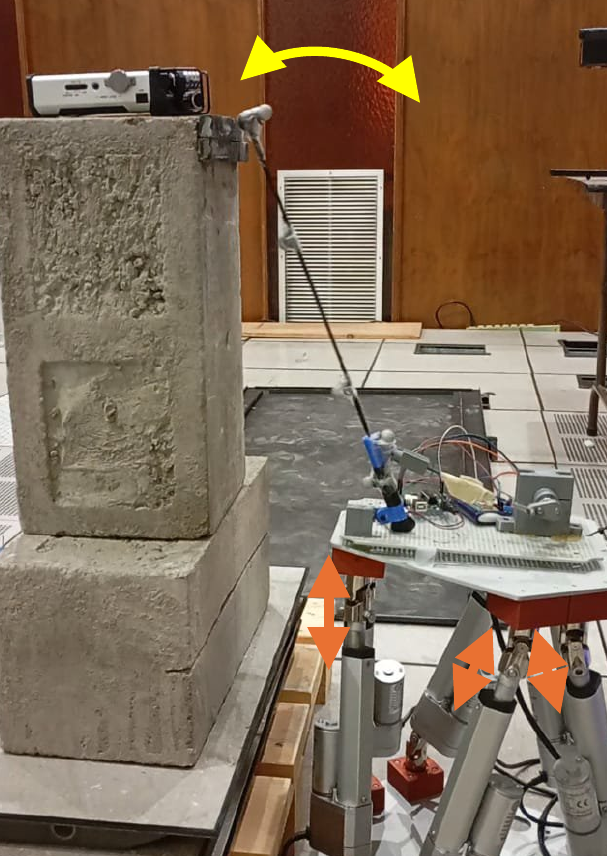}
    \caption{Hammer-specimen interaction based on the movement of the haxapod platform}
    \label{fig:oscillation}
\end{figure}

In Figure~\ref{fig:oscillation} shows that, as the degrees of freedom increase, the platform exhibits displacement in all three angles: pitch, roll, and yaw. This movement reduces the impact, or the energy with which the hammer would strike in the absence of disturbances, resulting in lower decibel sound peaks at certain time intervals.

Under ideal conditions specifically in Section \ref{zerocondS}, when there is no disturbance or vibration during data collection, the score is quite satisfactory, the metrics evaluated for samples without perturbation are calculated as 1 in their performance. However, when controlled vibrations were introduced and incrementally increased, a discernible degradation in performance was observed, evidences the impact on the score. 

This behavior is expected and consistent with the reduction in impact energy and sound peaks described above, as the movement of the platform alters the interaction between the hammer and the specimen over time. Although the decline in scoring metrics was not substantial (remaining above of 0.5), the trend indicates a measurable influence of vibrational interference on the system's accuracy. This observation suggests that while the system maintains robustness under mild perturbations, its effectiveness diminishes proportionally with increasing vibrational amplitude. Further analysis is required to determine the threshold at which such disturbances significantly impair functionality.

\subsection{Acoustic energy processing performance}


The previous section showed how the accuracy of the results changes based on the levels of disturbance that were introduced. In order to reduce error, the method implemented in Section~\ref{energyMethod} uses a peak selection approach based on the concentration of energy from my hits or hammer sounds when evaluating healthy and unhealthy surfaces to filter the most relevant acoustic events in tile inspection using NDT. This technique operates through two fundamental mechanisms: (1) elimination of low-amplitude and high-amplitude artifacts associated with environmental vibrations by optimal thresholds, and (2) preservation of acoustic characteristic of mechanical impact. Optimal threshold selection captures the complete acoustic response without including background disturbance.

\begin{figure}[htbp]
\centering 
\begin{subfigure}[b]{0.335\linewidth}
\includegraphics[width=0.99\textwidth]{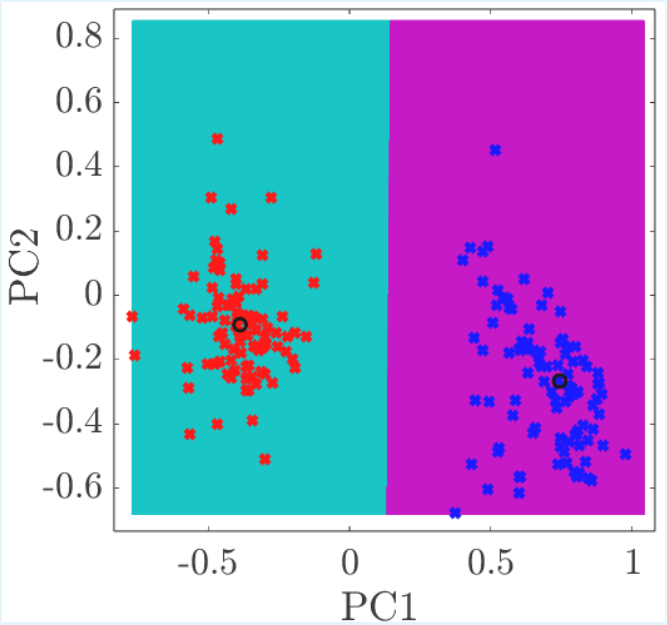}
\label{fig:a} \subcaption{} \end{subfigure} \hspace{1mm}
\begin{subfigure}[b]{0.295\linewidth}
\includegraphics[width=0.99\textwidth]{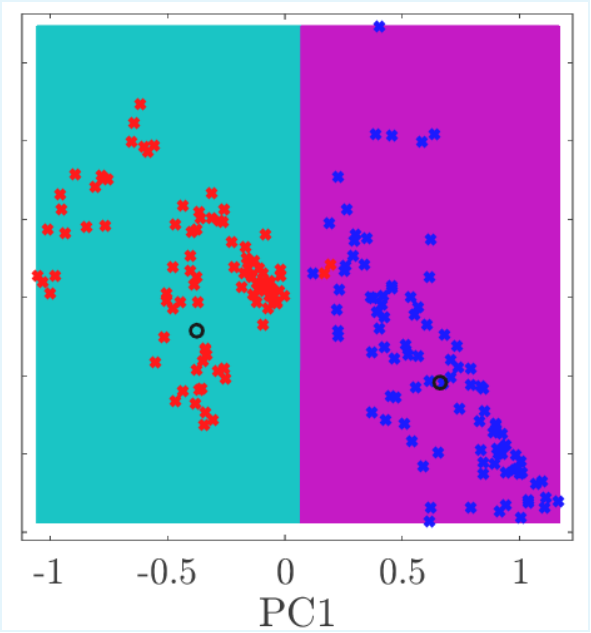}
\label{fig:b} \subcaption{} \end{subfigure} \hspace{-4mm}
\begin{subfigure}[b]{0.33\linewidth}
\includegraphics[width=0.99\textwidth]{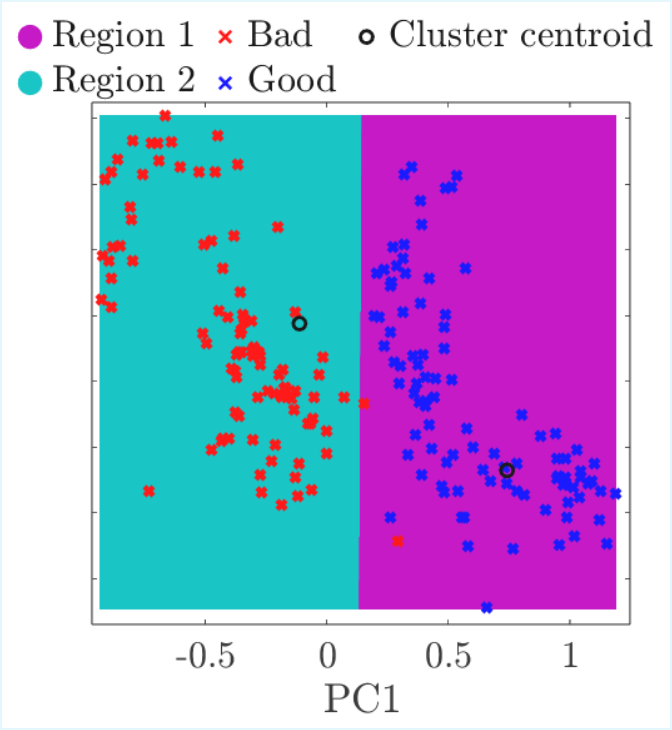}
\label{fig:b} \subcaption{} \end{subfigure}
\caption{PCA (a) 1 deg; (b) 3 deg; (c) 5 deg} \label{fig:resultTotalPCA22}
\end{figure} 

Figure~\ref{fig:resultTotalPCA22} shows that all cases remain within their respective PCA regions; however, there is still a level of dispersion within each region compared to the results obtained before applying the energy-based method. This is due to the fact that the method preserves the characteristics of minimal variability, which results in a clearer, but slightly more dispersed, representation within the principal component space without causing large classification errors between regions, as you can see in Table~\ref{tabC:divisionwithVibrationv2}.

\begin{table}[htbp]
    \centering \footnotesize 
    \caption{Results of the performance evaluation for training a testing.}
    \begin{tabular}{|c|c|c|c|c|c|c|c|c|}
    \hline
                   & \multicolumn{2}{c|}{0 deg} & \multicolumn{2}{c|}{1 deg} & \multicolumn{2}{c|}{3 deg} & \multicolumn{2}{c|}{5 deg} \\  \hline 
                   Condition & Good & Bad & Good & Bad & Good & Bad & Good & Bad \\  \hline
    Total Taps           & 38 & 39 & 96 & 97 & 96 & 97 & 96 & 97 \\ \hline
    Correctly classified & 38 & 39 & 96 & 97 & 96 & 95 & 96 & 95 \\ \hline
    Incorrectly classified & 0 & 0 & 0 & 0 & 0 & 2 & 0 & 2 \\ \hline
    \end{tabular}
    \label{tabC:divisionwithVibrationv2}
\end{table}


\subsection{Global performance metrics}

As shown in Table~\ref{tab:deggeneralScoreGlobalfinal}, the energy-based method demonstrated robustness at all vibration levels, consistently reaching 1.00 for the first vibration level and maintaining a net accuracy value above 0.98 even at the highest vibration level. 
The results provide a quantitative relationship between vibration amplitude and classification degradation, demonstrating that dynamic perturbations systematically distort acoustic feature distributions.

The proposed energy-based method effectively compensates for vibration-induced signal distortion by selectively preserving physically meaningful acoustic events while suppressing motion-related artifacts.
The performance of the proposed energy-based detection method was evaluated and compared with a reference detection method (without energy method) under different vibration levels (0°, 1°, 3°, and 5°). The evaluation was carried out using standard classification metrics, where the overall metric is Accuracy, and performance clearly improves, especially for the last two cases, which are the most critical. 

\input{Table/tableGlobal}

\begin{figure}[htbp]
    \centering 
    \includegraphics[width=0.99\linewidth]{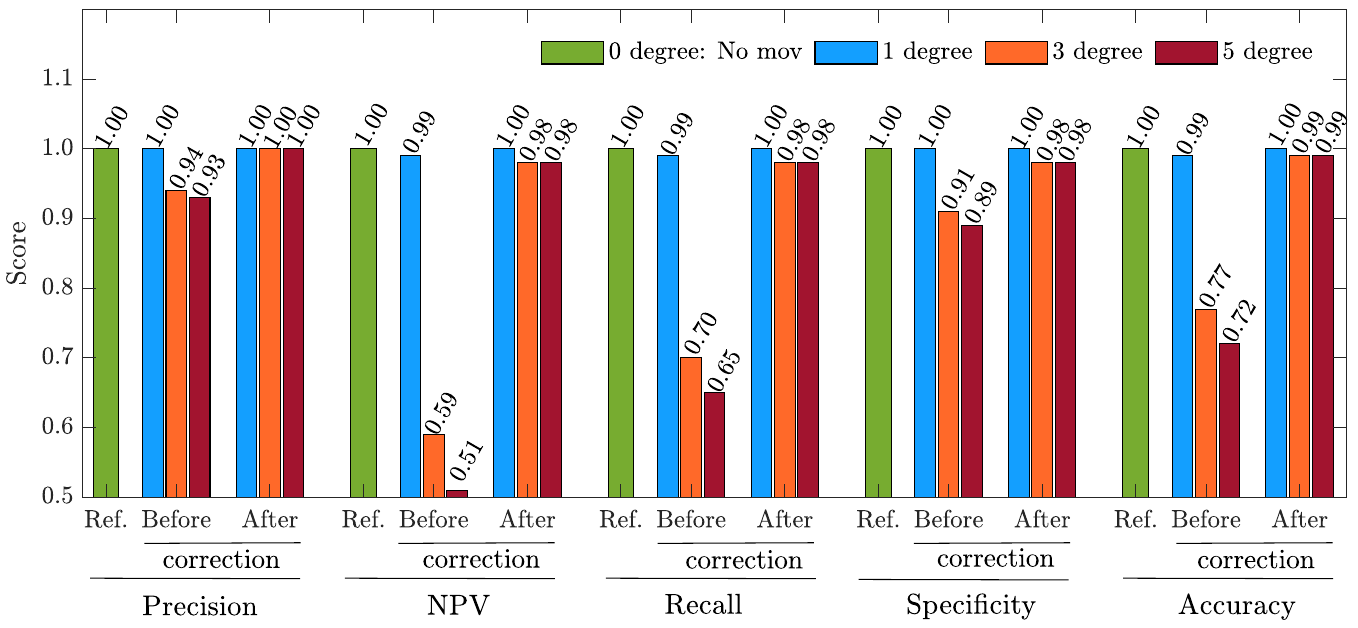}
    \caption{Overall performance at 0 degrees (no movement), 1 degree, 3 degrees, and 5 degrees of perturbation}
    \label{fig:ScorePlot135degDinal}
\end{figure}

Figure~\ref{fig:ScorePlot135degDinal} also shows that under ideal conditions (0° vibration), both methods worked perfectly, achieving the maximum score. However, as vibration increased, the reference method showed a notable decrease in performance, while the proposed energy method maintained Accuracy above 0.98.

The diagnosis of the structural condition of facades through the evaluation of tiles with the tap testing system showed a significant improvement after the implementation of the methodological corrections proposed in this study. However, it should be noted that the experiments were conducted under controlled laboratory conditions, with predefined UAV trajectories. This experimental setup could differ from real operating conditions, where various environmental factors such as wind gusts, thermal variations and the nonlinear dynamics of the UAVs could influence the accuracy of the measurements.

%% file: Table/tableGeneralResult.tex
\begin{table}[htbp]
    \centering \footnotesize 
    \caption{Results of the performance evaluation for each vibration level.}
    \begin{tabular}{|m{2.6cm}|m{1.4cm}|m{1.4cm}|m{1.4cm}|m{1.4cm}|}
    \hline
    \textbf{Condition} & No Osc & 1 deg & 3 deg & 5 deg \\  \hline 
    \textbf{Precision} & 1.00 & 1.00 & 0.94 & 0.93 \\
    \hline
    \textbf{NPV} & 1.00 & 0.99 & 0.59 & 0.51 \\    
    \hline
    \textbf{Recall}  & 1.00 & 0.99 & 0.70 & 0.65 \\    
    \hline
    \textbf{Specificity} & 1.00 & 1.00 & 0.91 & 0.89 \\    
    \hline
    \textbf{Accuracy}  & 1.00 & 0.99 & 0.77 & 0.72 \\
    \hline
    \end{tabular}
    \label{tab:generalScore1deg}
\end{table}

%% file: Table/tableGlobal.tex
\begin{table}[ht]
    \centering \footnotesize 
    \caption{Overall performance of metrics for each case.}
    \begin{tabular}{|ccccccc|}
    \hline
                    \textbf{Vibration level} & \textbf{Case} & \textbf{Precision} & \textbf{NPV} & \textbf{Recall} & \textbf{Specificity} & \textbf{Accuracy} \\  \hline 
    0 deg & Base & 1.00 & 1.00  & 1.00 & 1.00 & 1.00 \\
    \hline
    
    \multirow{2}{*}{1 deg} & w/o energy method & 1.00 & 0.99  & 0.99 & 1.00 & 0.99 \\  
     & Energy method & 1.00 & 1.00  & 1.00 & 1.00 & 1.00 \\
    \hline  
    
    \multirow{2}{*}{3 deg} & w/o energy method & 0.94 & 0.59  & 0.70 & 0.91 & 0.77 \\  
     & Energy method & 1.00 & 0.98  & 0.98 & 0.98 & 0.99 \\
    \hline  
    
    \multirow{2}{*}{5 deg} & w/o energy method & 0.93 & 0.51  & 0.65 & 0.89 & 0.72 \\  
     & Energy method & 1.00 & 0.98  & 0.98 & 0.98 & 0.99\\
    \hline
\end{tabular}
\label{tab:deggeneralScoreGlobalfinal}
\end{table}

%% file: Sec7.tex

\section{Conclusions and Future work}


This work establishes a controlled validation framework that bridges the gap between laboratory acoustic tap-testing and UAV-based facade inspection. The results demonstrate that UAV-induced perturbations significantly degrade classification performance, but this effect can be systematically mitigated using an energy-based signal processing approach. The study provides a reproducible quantification of motion-induced acoustic degradation and its correction, forming a necessary pre-field validation stage for robotic façade inspection systems.
The method of this research automates the traditional hammer test procedure and mounts the hardware system on a UAV platform and collects the acoustic data from the hammer and then uses PCA analysis to classify the different quality specimens into different clusters. Subsequently, the specimen clusters and their principal values are used to generate classifications and are projected onto a 2D principal component space. This allows us to classify the condition of the facade tiles with different levels of vibration, i.e., whether they are intact or show any detachment, especially in tall buildings.

These findings highlight the sensitivity of the system to mechanical perturbations. While the baseline condition produces tightly grouped clusters, the introduction of vibration leads to a measurable shift in data distribution. This effect, though systematic, does not entirely obscure classification boundaries, indicating that the system retains some robustness under controlled disturbances. 
As a measure to reduce errors, the integration of the spectral energy-based threshold serves to mitigate such interference and improve damage detection in dynamic and noisy environments. This method improves detection robustness by adaptively refining the signal using Parseval-derived energy estimates, effectively isolating actual acoustic events from vibration noise. Therefore, the proposed energy-based framework not only compensates for environmental variability but also provides a scalable approach to ecoacoustic monitoring.




In addition to reducing the error by energy method, the implementation of a stabilization physical system to minimize the disturbances affecting the drone is proposed as future work. Also, this work will extend the framework to higher-complexity models once the underlying physical behavior is fully characterized.
Finally, field experiments are planned to validate the assumptions and considerations made in the laboratory.




\section{Acknowledgement}

The authors would like to thank the U.S. Department of Energy’s Office of Environmental (DOE-EM) for their financial support and guidance. The research was conducted under direct support from DOE-EM under a DOE-FIU Cooperative Agreement (Contract No. DE-EM0005213). The authors are also grateful to Kira Kampschmidt and Michael Carl for their assistance with this research.

%% file: sample.bib
@techreport{mason2016tap,
  title={Tap testing hammer using unmanned aerial systems (UASs)},
  author={Mason, JaMein DeShon and Ayorinde, Emmanuel Temiloluwa and Mascarenas, David Dennis and Moreu, Fernando},
  year={2016},
  institution={Los Alamos National Lab.(LANL), Los Alamos, NM (United States)}
}

@article{buxton2018efficacy,
  title={Efficacy of extracting indices from large-scale acoustic recordings to monitor biodiversity},
  author={Buxton, Rachel T and McKenna, Megan F and Clapp, Maddie and others},
  journal={Conservation Biology},
  volume={32},
  number={5},
  pages={1174--1184},
  year={2018}
}

@article{sanchez2024exploring,
  title={Exploring fish choruses: patterns revealed through PCA computed from daily spectrograms},
  author={S{\'a}nchez-Gendriz, Ignacio and Luna-Naranjo, David and Guedes, Luiz Affonso and L{\'o}pez, Jos{\'e} D and Padovese, Linilson R},
  journal={Frontiers in Antennas and Propagation},
  volume={2},
  pages={1400382},
  year={2024},
  publisher={Frontiers Media SA}
}

@article{sanchez2021signal,
  title={Signal processing basics applied to ecoacoustics},
  author={Sanchez-Gendriz, Ignacio},
  journal={Ecological Informatics},
  volume={66},
  pages={101445},
  year={2021},
  publisher={Elsevier}
}

@article{wang2025road,
  title={Road disturbance drives a more simplified soundscape in temperate forests revealed by deep learning and acoustics indices},
  author={Wang, Shizheng and Duan, Yuxuan and Cao, Ranxing and Feng, Jiawei and Ge, Jianping and Wang, Tianming},
  journal={Biological Conservation},
  volume={306},
  pages={111115},
  year={2025},
  publisher={Elsevier}
}

@article{garg2020measuring,
  title={Measuring transverse displacements using unmanned aerial systems laser Doppler vibrometer (UAS-LDV): Development and field validation},
  author={Garg, Piyush and Nasimi, Roya and Ozdagli, Ali and Zhang, Su and Mascarenas, David Dennis Lee and Reda Taha, Mahmoud and Moreu, Fernando},
  journal={Sensors},
  volume={20},
  number={21},
  pages={6051},
  year={2020},
  publisher={MDPI}
}

@misc{viconmodel,
author = {VICON},
title = {Vicon camera valkyrie},
howpublished = {11 Nov. 2017 [On line]. Available: \url{https://www.vicon.com/hardware/cameras/valkyrie/} [accessed on 17 July 2025]}
}

@misc{dronemodel,
author = {DJI},
title = {Drone Matrice 600 Pro},
howpublished = {11 Nov. 2014 [On line]. Available: \url{https://www.dji.com/support/product/matrice600-pro} [accessed on 17 July 2025]}
}

@article{nemati2023biomimetic,
  title={Biomimetic investigation of the impact of the ear canal on the acoustic field sensitivity of aye-ayes},
  author={Nemati, Hamidreza and Dehghan-Niri, Ehsan},
  journal={Applied Acoustics},
  volume={202},
  pages={109171},
  year={2023},
  publisher={Elsevier}
}

@inproceedings{sukvichai2017design,
  title={Design of a double-propellers wall-climbing robot},
  author={Sukvichai, Kanjanapan and Maolanon, Pruttapon and Songkrasin, Konlayut},
  booktitle={2017 ieee international conference on robotics and biomimetics (ROBIO)},
  pages={239--245},
  year={2017},
  organization={IEEE}
}

@inproceedings{masurkar2024enhancing,
  title={Enhancing biomimetic design of tap scanning sensors through high-resolution thermal camera-based behavioral studies},
  author={Masurkar, Nihar and Nemati, Hamidreza and Dehghan-Niri, Ehsan},
  booktitle={Bioinspiration, Biomimetics, and Bioreplication XIV},
  volume={12944},
  pages={93--99},
  year={2024},
  organization={SPIE}
}

@article{nemati2025investigating,
  title={Investigating Acoustic-based Foraging Behavior in Aye-ayes (Daubentonia madagascariensis) Through Infrared Thermography},
  author={Nemati, Hamidreza and Masurkar, Nihar and Dehghan-Niri, Ehsan},
  journal={International Journal of Primatology},
  pages={1--22},
  year={2025},
  publisher={Springer}
}

@article{huang2023development,
  title={Development of a variable-frequency hammering method using acoustic features for damage-type identification},
  author={Huang, Xi and Huang, Huang and Wu, Zhishen},
  journal={Applied Sciences},
  volume={13},
  number={3},
  pages={1329},
  year={2023},
  publisher={MDPI}
}

@article{shoda2024defect,
  title={Defect detection with ego-noise reduction based on multimodal information in UAV hammering inspection},
  author={Shoda, Koki and Louhi Kasahara, Jun Younes and Asama, Hajime and An, Qi and Yamashita, Atsushi},
  journal={Advanced Robotics},
  volume={38},
  number={17},
  pages={1218--1230},
  year={2024},
  publisher={Taylor \& Francis}
}

@article{song2024tunnel,
  title={Tunnel lining quality detection technology based on impulse echo acoustic method from fine management perspective},
  author={Song, Jingjing and Feng, Yuan and Huang, Botai},
  journal={Wireless Networks},
  pages={1--12},
  year={2024},
  publisher={Springer}
}

@article{nishimura2024propeller,
  title={Propeller-type wall-climbing robot for visual and hammering inspection of concrete surfaces},
  author={Nishimura, Yuki and Mochiyama, Hiromi and Yamaguchi, Tomoyuki},
  journal={IEEE Access},
  year={2024},
  publisher={IEEE}
}

@article{abdi2010principal,
  title={Principal component analysis},
  author={Abdi, Herv{\'e} and Williams, Lynne J},
  journal={Wiley interdisciplinary reviews: computational statistics},
  volume={2},
  number={4},
  pages={433--459},
  year={2010},
  publisher={Wiley Online Library}
}

@misc{FlyskyRemote,
author = {Flysky},
title = {Flysky FS-i6X 6-10(Default 6)CH 2.4GHz AFHDS RC Transmitter w/ FS-iA6B Receiver},
howpublished = {11 Nov. 2019 [On line]. Available: \url{https://www.amazon.com/Flysky-FS-i6X-Transmitter-FS-iA6B-Receiver/dp/B0744DPPL8/ref=sr_1_1?crid=2QVJIWXH1ZZ05&keywords=15.+Flysky+FS-i6X+6-10%28Default+6%29CH+2.4GHz+AFHDS+RC+Transmitter+w%2F+FS-iA6B+Receiver.&qid=1644535926&sprefix=15.+flysky+fs-i6x+6-10+default+6+ch+2.4ghz+afhds+rc+transmitter+w%2F+fs-ia6b+receiver.+%2Caps%2C693&sr=8-1} [accessed on 5 June 2025]}
}

@misc{microphonePCM,
author = {TASCAM},
title = {DR-44WL Portable Handheld Recorder with Wi-Fi},
howpublished = {20 Oct. 2022. [On line]. Available: \url{https://tascam.com/us/product/dr-44wl} [accessed on 4 June 2025]}
}

@misc{vibrationHexapod,
author = {Serhan Argun},
title = {Vibration and Oscillation in Stewart Platform},
howpublished = {27 Sep. 2024. [On line]. Available: \url{https://acrome.net/post/vibration-and-oscillation-in-stewart-platform} [accessed on 4 June 2025]}
}

@mastersthesis{adhikari2023usage,
  title={Usage of Drone for Building Facade Inspection},
  author={Adhikari, Sahara},
  year={2023},
  school={Marquette University}
}

@book{goodfellow2016deep,
  title={Deep learning},
  author={Goodfellow, Ian and Bengio, Yoshua and Courville, Aaron and Bengio, Yoshua},
  volume={1},
  number={2},
  year={2016},
  publisher={MIT press Cambridge}
}

@book{brunton2022data,
  title={Data-driven science and engineering: Machine learning, dynamical systems, and control},
  author={Brunton, Steven L and Kutz, J Nathan},
  year={2022},
  publisher={Cambridge University Press}
}

@article{rakha2018review,
  title={Review of Unmanned Aerial System (UAS) applications in the built environment: Towards automated building inspection procedures using drones},
  author={Rakha, Tarek and Gorodetsky, Alice},
  journal={Automation in construction},
  volume={93},
  pages={252--264},
  year={2018},
  publisher={Elsevier}
}

@article{dias2021critical,
  title={Critical analysis about emerging technologies for building’s fa{\c{c}}ade inspection},
  author={Dias, Il{\'\i}dio S and Flores-Colen, In{\^e}s and Silva, Ana},
  journal={Buildings},
  volume={11},
  number={2},
  pages={53},
  year={2021},
  publisher={MDPI}
}

@article{laofor2012defect,
  title={Defect detection and quantification system to support subjective visual quality inspection via a digital image processing: A tiling work case study},
  author={Laofor, Chollada and Peansupap, Vachara},
  journal={Automation in Construction},
  volume={24},
  pages={160--174},
  year={2012},
  publisher={Elsevier}
}

@article{hassani2023systematic,
  title={A systematic review of advanced sensor technologies for non-destructive testing and structural health monitoring},
  author={Hassani, Sahar and Dackermann, Ulrike},
  journal={Sensors},
  volume={23},
  number={4},
  pages={2204},
  year={2023},
  publisher={MDPI}
}

@article{avci2021review,
  title={A review of vibration-based damage detection in civil structures: From traditional methods to Machine Learning and Deep Learning applications},
  author={Avci, Onur and Abdeljaber, Osama and Kiranyaz, Serkan and Hussein, Mohammed and Gabbouj, Moncef and Inman, Daniel J},
  journal={Mechanical systems and signal processing},
  volume={147},
  pages={107077},
  year={2021},
  publisher={Elsevier}
}

@article{bodie2019omnidirectional,
  title={An omnidirectional aerial manipulation platform for contact-based inspection},
  author={Bodie, Karen and Brunner, Maximilian and Pantic, Michael and Walser, Stefan and Pf{\"a}ndler, Patrick and Angst, Ueli and Siegwart, Roland and Nieto, Juan},
  journal={arXiv preprint arXiv:1905.03502},
  year={2019}}

@article{zhang2017force,
  title={A force-sensing system on legs for biomimetic hexapod robots interacting with unstructured terrain},
  author={Zhang, He and Wu, Rui and Li, Changle and Zang, Xizhe and Zhang, Xuehe and Jin, Hongzhe and Zhao, Jie},
  journal={Sensors},
  volume={17},
  number={7},
  pages={1514},
  year={2017},
  publisher={MDPI}}

@inproceedings{martonka2014hexapod,
  title={Hexapod: The Platform with 6DOF},
  author={Martonka, R and Fliegel, V},
  booktitle={Modern Methods of Construction Design: Proceedings of ICMD 2013},
  pages={133--138},
  year={2014},
  organization={Springer}}

@article{min2021port,
  title={Port structure inspection based on 6-DOF displacement estimation combined with homography formulation and genetic algorithm},
  author={Min, Jiyoung and Bang, Yuseok and Bang, Hyuntae and Jeon, Haemin},
  journal={Applied Sciences},
  volume={11},
  number={14},
  pages={6470},
  year={2021},
  publisher={MDPI}
}

@article{burkus2022mechanical,
  title={Mechanical design and a novel structural optimization approach for hexapod walking robots},
  author={Burkus, Ervin and Odry, {\'A}kos and Awrejcewicz, Jan and Kecsk{\'e}s, Istv{\'a}n and Odry, P{\'e}ter},
  journal={Machines},
  volume={10},
  number={6},
  pages={466},
  year={2022},
  publisher={MDPI}
}

@article{teixeira2021intelligent,
  title={An intelligent hexapod robot for inspection of airframe components oriented by deep learning technique},
  author={Teixeira Vivaldini, Kelen C and Franco Barbosa, Gustavo and Santos, Igor Araujo Dias and Kim, Pedro HC and McMichael, Grayson and Guerra-Zubiaga, David A},
  journal={Journal of the Brazilian Society of Mechanical Sciences and Engineering},
  volume={43},
  pages={1--15},
  year={2021},
  publisher={Springer}
}

@inproceedings{silva2024automated,
  title={Automated facade inspection: Application and challenge in using Artificial Intelligence for construction defect recognition},
  author={Silva, Alisson Souza and MELO, Roseneia Rodrigues Santos and COSTA, Dayana Bastos},
  booktitle={XX International Conference on Building Pathology and Constructions Repair},
  pages={705--718},
  year={2024}
}

@inproceedings{peng2017unmanned,
  title={Unmanned Aerial Vehicle for infrastructure inspection with image processing for quantification of measurement and formation of facade map},
  author={Peng, KC and Feng, L and Hsieh, YC and Yang, TH and Hsiung, SH and Tsai, YD and Kuo, C},
  booktitle={2017 international conference on applied system innovation (ICASI)},
  pages={1969--1972},
  year={2017},
  organization={IEEE}}

@article{falorca2021facade,
  title={Facade inspections with drones--theoretical analysis and exploratory tests},
  author={Falorca, Jorge Furtado and Lanzinha, Jo{\~a}o Carlos Gon{\c{c}}alves},
  journal={International Journal of Building Pathology and Adaptation},
  volume={39},
  number={2},
  pages={235--258},
  year={2021},
  publisher={Emerald Publishing Limited}
}

@inproceedings{chen2019opportunities,
  title={Opportunities for applying camera-equipped drones towards performance inspections of building facades},
  author={Chen, Kaiwen and Reichard, Georg and Xu, Xin},
  booktitle={ASCE International Conference on Computing in Civil Engineering 2019},
  pages={113--120},
  year={2019},
  organization={American Society of Civil Engineers Reston, VA}
}

@inproceedings{ruiz2021unmanned,
  title={Unmanned Aerial Vehicles and Digital Image Processing with Deep Learning for the Detection of Pathological Manifestations on Facades},
  author={Ruiz, Ramiro Daniel Ballesteros and Lordsleem J{\'u}nior, Alberto Casado and Fernandes, Bruno Jos{\'e} Torres and Oliveira, S{\'e}rgio Campello},
  booktitle={Proceedings of the 18th International Conference on Computing in Civil and Building Engineering: ICCCBE 2020},
  pages={1099--1112},
  year={2021},
  organization={Springer}
}

@article{silva2022stewart,
  title={Stewart platform motion control automation with industrial resources to perform cycloidal and oceanic wave trajectories},
  author={Silva, Diego and Garrido, Julio and Riveiro, Enrique},
  journal={Machines},
  volume={10},
  number={8},
  pages={711},
  year={2022},
  publisher={MDPI}
}

@article{lin2025true,
  title={True 3D thermal inspection of buildings using multimodal UAV images},
  author={Lin, Dong and Yang, Na and Miao, Qi and Cui, Xiaojie and Xu, Dinggen},
  journal={Journal of Building Engineering},
  volume={100},
  pages={111806},
  year={2025},
  publisher={Elsevier}
}

@article{zhang2024integrated,
  title={An Integrated Method Using a Convolutional Autoencoder, Thresholding Techniques, and a Residual Network for Anomaly Detection on Heritage Roof Surfaces},
  author={Zhang, Yongcheng and Kong, Liulin and Antwi-Afari, Maxwell Fordjour and Zhang, Qingzhi},
  journal={Buildings},
  volume={14},
  number={9},
  pages={2828},
  year={2024},
  publisher={MDPI}
}

@article{mayya2025triple,
  title={Triple-stage crack detection in stone masonry using YOLO-ensemble, MobileNetV2U-net, and spectral clustering},
  author={Mayya, Ali Mahmoud and Alkayem, Nizar Faisal},
  journal={Automation in Construction},
  volume={172},
  pages={106045},
  year={2025},
  publisher={Elsevier}
}

@article{karimi2024automated,
  title={Automated surface crack detection in historical constructions with various materials using deep learning-based YOLO network},
  author={Karimi, Narges and Mishra, Mayank and Louren{\c{c}}o, Paulo B},
  journal={International Journal of Architectural Heritage},
  pages={1--17},
  year={2024},
  publisher={Taylor \& Francis}
}

@article{dong2025new,
  title={A New Method for Rapid Detection of Surface Defects on Complex Textured Tiles},
  author={Dong, Guanping and Wang, Yuanzhi and Liu, Sai and Wu, Nanshou and Kong, Xiangyu and Chen, Xiangyang and Wang, Zixi},
  journal={Journal of Nondestructive Evaluation},
  volume={44},
  number={1},
  pages={1--17},
  year={2025},
  publisher={Springer}
}

@inproceedings{ichikawa2017uav,
  title={UAV with manipulator for bridge inspection—Hammering system for mounting to UAV},
  author={Ichikawa, Akihiko and Abe, Yuki and Ikeda, Takahiro and Ohara, Kenichi and Kishikawa, Jyunpei and Ashizawa, Satoshi and Oomichi, Takeo and Okino, Akihisa and Fukuda, Toshio},
  booktitle={2017 IEEE/SICE International Symposium on System Integration (SII)},
  pages={775--780},
  year={2017},
  organization={IEEE}
}

@article{karimi2024deep,
  title={Deep learning-based automated tile defect detection system for Portuguese cultural heritage buildings},
  author={Karimi, Narges and Mishra, Mayank and Louren{\c{c}}o, Paulo B},
  journal={Journal of Cultural Heritage},
  volume={68},
  pages={86--98},
  year={2024},
  publisher={Elsevier}
}

@article{nishimura2022automated,
  title={Automated hammering inspection system with multi-copter type mobile robot for concrete structures},
  author={Nishimura, Yuki and Takahashi, Shuki and Mochiyama, Hiromi and Yamaguchi, Tomoyuki},
  journal={IEEE Robotics and Automation Letters},
  volume={7},
  number={4},
  pages={9993--10000},
  year={2022},
  publisher={IEEE}
}

@article{nemati2024bio,
  title={Bio-inspired robotic tap testing: An innovative approach for nondestructive testing of wooden structures},
  author={Nemati, Hamidreza and Dehghan-Niri, Ehsan},
  year={2024}
}

@article{nasimi2021crack,
  title={Crack detection using tap-testing and machine learning techniques to prevent potential rockfall incidents},
  author={Nasimi, Roya and Moreu, Fernando and Stormont, John},
  journal={Engineering Research Express},
  volume={3},
  number={4},
  pages={045050},
  year={2021},
  publisher={IOP Publishing}
}

@article{nasimi2022use,
  title={Use of remote structural tap testing devices deployed via ground vehicle for health monitoring of transportation infrastructure},
  author={Nasimi, Roya and Atcitty, Solomon and Thompson, Dominic and Murillo, Joshua and Ball, Marlan and Stormont, John and Moreu, Fernando},
  journal={Sensors},
  volume={22},
  number={4},
  pages={1458},
  year={2022},
  publisher={MDPI}
}

@article{moreu2018remote,
  title={Remote railroad bridge structural tap testing using aerial robots},
  author={Moreu, F and Ayorinde, E and Mason, J and Farrar, C and Mascarenas, D},
  journal={International Journal of Intelligent Robotics and Applications},
  volume={2},
  number={1},
  pages={67--80},
  year={2018},
  publisher={Springer}
}
